\documentclass[a4paper]{report}
\usepackage[T1]{fontenc}
\usepackage{mynm}
\usepackage[CaptionBefore]{fltpage}
\usepackage{titlesec}
\usepackage{mypyg}
\usepackage{etoolbox}
\patchcmd{\thebibliography}{\chapter*}{\section*}{}{}

\titlespacing*{\chapter}{0pt}{-20pt}{3ex}
\titleformat{\chapter}{\normalfont\LARGE\bfseries}{Supplementary Note \thechapter.}{10pt}{}

\DefineShortVerb{\|}
\fvsettext

\begin{document}
\thispagestyle{empty}
$ $
\vspace{6cm}

{\centering
{\huge\textbf{BigDataViewer: Interactive Visualization and\\[2mm] Image Processing for Terabyte Data Sets}}\\[1cm]
Tobias Pietzsch$^\text{1}$, Stephan Saalfeld$^\text{2}$, Stephan Preibisch$^\text{1,2,3}$, Pavel Tomancak$^\text{1}$\\[2cm]
$^\text{1}$Max Planck Institute of Molecular Cell Biology and Genetics, 01307 Dresden, Germany\\[3mm]
$^\text{2}$Janelia Farm Research Campus, Howard Hughes Medical Institute, Ashburn, VA 20147, USA\\[3mm]
$^\text{3}$Department of Anatomy and Structural Biology, Gruss Lipper Biophotonics Center,\\ Albert Einstein College of Medicine, Bronx, NY 10461, USA\\

\vspace{8.7cm}
Correspondence should be addressed to: pietzsch@mpi-cbg.de and tomancak@mpi-cbg.de\\
}

\newpage
\noindent
The increasingly popular light sheet microscopy
techniques generate very large 3D time-lapse recordings of living biological specimen\cite{SPIM}.
The necessity to make large volumetric datasets available for interactive
visualization and analysis has been widely recognized\cite{huszal12}.
However, existing solutions build on dedicated servers to generate virtual slices that are transferred to the client applications, practically leading to insufficient frame rates (less than 10 frames per second) for truly interactive experience.
An easily accessible open source solution for interactive arbitrary virtual re-slicing of very large volumes and time series of volumes has yet been missing.

We fill this gap with BigDataViewer (BDV), a Fiji plugin~\cite{fiji} to interactively navigate and visualize
large image sequences from both local and remote data sources.  The client software renders an
arbitrarily oriented virtual slice through global 3D coordinate space (\textbf{Fig.~\ref{fig:fig1}a}). Individual image stacks, each representing a view of a multi-view Selective Plane Illumination Microscopy (SPIM) data set, can be displayed independently or as a color-coded composite
(\textbf{Fig.~\ref{fig:fig1}a}). Brightness and color can be adjusted for each view separately. The
viewer allows free translation, rotation, and zoom for image stacks and moving between
timepoints. Thus, multi-terabyte recordings can be navigated smoothly (\textbf{Supplementary Video 1}).

We achieve this performance on very large data by an efficient client-side renderer, and an intelligent
loading and caching scheme. To render any virtual slice, only a small
fraction of the image data is relevant and gets loaded into memory.
The navigation is further accelerated by caching in memory
recently visited locations. If images are available at multiple scales, only the most relevant scales for display are requested. This avoids aliasing artifacts at
zoomed-out views and facilitates interactive browsing: low-resolution data are
loaded rapidly, providing immediate feedback, while high-resolution detail is
filled in subsequently (\textbf{Supp\-lementary Note~\supplRenderingNumber}).

For large time series, we developed a custom HDF5 based
data format that is optimized for fast random data access
at various scales (\textbf{Supp\-lementary Note \supplFileFormatNumber}). Each image is stored as a chunked multi-dimensional array
at successively reduced resolutions.
We build on HDF5 as an established portable data format that provides efficient
input and output, supports unlimited file sizes and has built-in and extensible
compression facilities. Metadata, such as SPIM view registrations, are
stored as XML. The proposed format integrates seamlessly with Fiji's plugins for SPIM image processing
allowing to control and visualize the
intermediate steps of the pipeline (\textbf{Fig.~\ref{fig:fig1}b-e}). We
provide Fiji plugins for reading and writing the format so that any image that
opens in Fiji can be stored as HDF5 and viewed with the BDV plugin (\textbf{Supp\-lementary Note \supplPluginsNumber}).

BigDataViewer has a modular architecture that separates data access, caching,
and visualization (\textbf{Supp\-lementary Note \supplArchitectureNumber}). We rely on the generic image
processing library ImgLib2~\cite{imglib2} to provide abstract interfaces between the
modules. This enables the data backend to hide implementation details such as
chunking and caching. The modular architecture makes it easy to add alternative
data backends for data sources other than the custom HDF5 format. For instance, BDV can access the online image services
CATMAID~\cite{catmaid} and OpenConnectome~\cite{ocp}. In this case it provides
re-slicing visualisation of the massive serial section electron microscopy datasets as an alternative to the fixed web-browser visualization (\textbf{Fig.~\ref{fig:fig1}f-i}). In addition, we
developed a web-service that allows to access HDF5 datasets remotely, enabling from within Fiji web browser like access to image data stored online.
In all cases, the data backend provides
cached virtualized access to image data, resulting in the illusion that all of the data are
present in memory.

Besides visualization, the virtualized access is a powerful way to
present remote and/or extremely large datasets for computation. In Fiji, we use
our framework to make raw images of the datasets available as virtual stacks
and can then run standard image processing tools on arbitrarily large images.
For example a SPIM time-lapse may be registered, fused, and deconvolved without
being locally stored on the processing computer. Moreover, it is
straightforward to programmatically access the pixel data using standard
ImgLib2 interfaces which means that existing code for filtering and segmentation
will work without modification (\textbf{Fig.~\ref{fig:fig1}k}). Similarly, the visualisation frontend can be
programmatically extended to display additional data, such as image processing
results or annotations (\textbf{Fig.~\ref{fig:fig1}l}).
Thus the BigDataViewer is a tool for visualisation and processing of the special case multi-view data from light sheet microscopy as well as a general solution for dealing with large terabyte sized datasets from any imaging modality.

\newpage
\paragraph*{Acknowledgements}\ \\
We thank Florian Jug for proofreading and helpful discussions.
We thank Evangelia Stamataki for datasets used in supplementary movies.
We thank Johannes Schindelin and Curtis Rueden for developing and maintaining the Fiji infrastructure.
SP was supported by MPI-CBG, HHMI and the Human Frontier Science Program (HFSP) Postdoctoral Fellowship.
PT and TP were supported by The European Research Council Community's Seventh Framework Program (FP7/2007-2013) grant agreement 260746.

\paragraph*{Author contributions}\ \\
TP and SS wrote the software.
SS and TP devised all algorithms.
TP and SP designed the file format and integrated with SPIM registration.
TP, SS, PT conceived the project and wrote the paper.

\bibliographystyle{ieeetr}

\begin{FPfigure}
  \includegraphics[width=\textwidth]{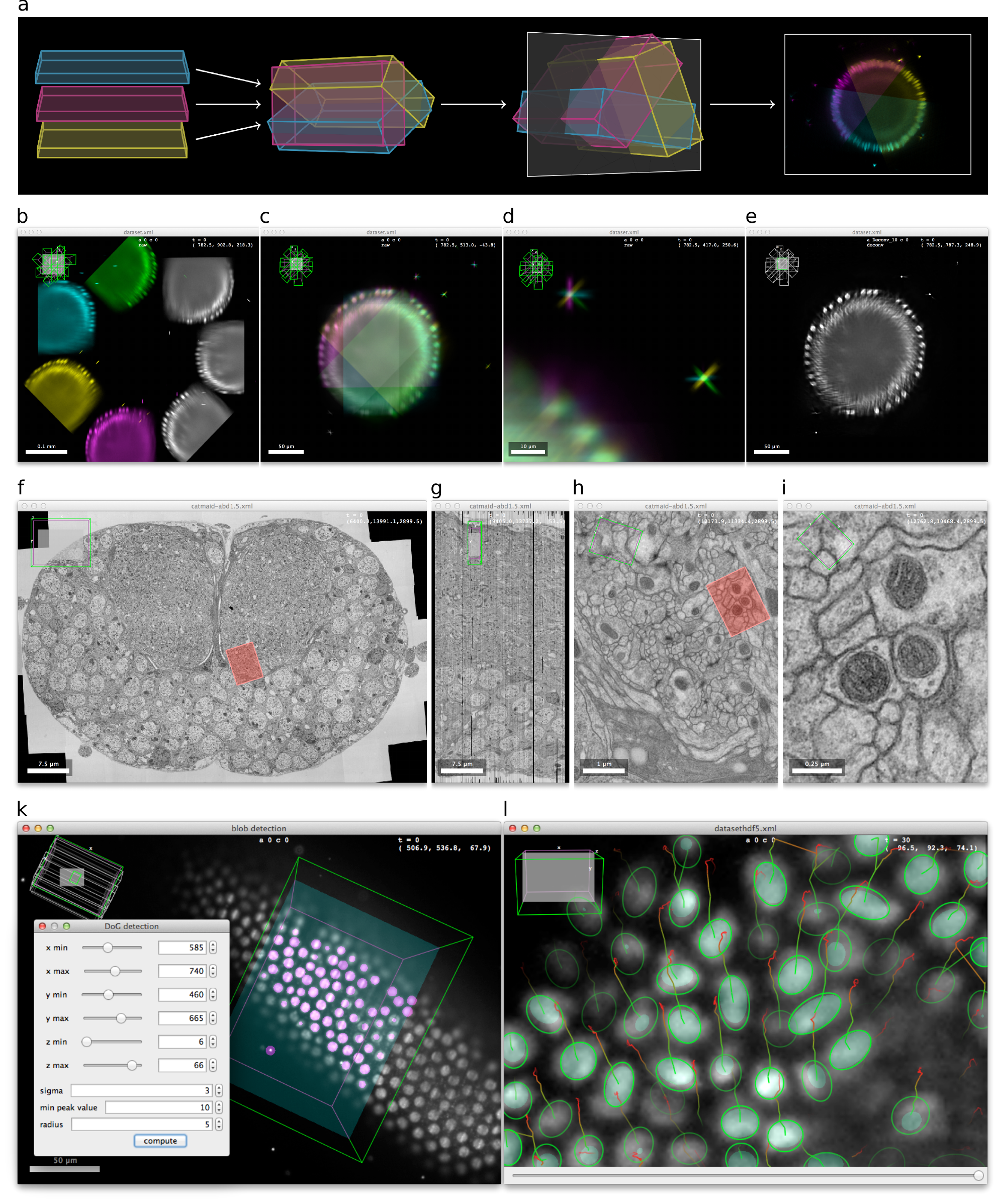}
  \caption{%
\textbf{(a)} Rendering algorithm.
  Source image volumes are transformed into a common global space.
  Then the global space into the current viewer frame which is aligned with the screen plane $z=0$.
  Pixels on this plane are rendered to the viewer canvas to produce a final image.
\textbf{(b-e)} Fiji's SPIM image processing integration.
  Individual angles of a light-sheet microscopy dataset can be interactively viewed before registration (\textbf{b}) and after registration (\textbf{c}).
  A zoomed-in view of fluorescent beads after registration is shown in (\textbf{d}).
  The results of a multiview-deconvolution\cite{deconv} and other processing steps can be incorporated into dataset and viewed in a common global space (\textbf{e}).
\textbf{(f-i)} Online data sources.
  An electron microscopy dataset of $1.5$ segments of the ventral nerve cord of
  a first-instar \textit{Drosophila} larva ($458$ sections,
      each section consisting of $\sim70$ overlapping
      image tiles, imaged at $4\text{nm}/\text{px}$ with $50\text{nm}$ section thickness)\cite{SaalfeldAl12}
  is available online through the web-browser based CATMAID viewer.
  BigDataViewer improves interactive visualization of the dataset with arbitrary 3D navigation using the same intuitive interface as for local data.
  (\textbf{f}) shows one z-section of the dataset, (\textbf{g}) shows a resliced x-section.
  (\textbf{h} and \textbf{i}) are zoomed-in views of the areas highlighted in (\textbf{f}) and (\textbf{h}).
\textbf{(k,l)} Extensibility and processing. Source volumes are accessible and processing results can be visualized via standard interfaces.
  As an example, a blob detection algorithm is run on a subvolume selected by the user, and results are visualized in BDV (\textbf{k}).
  Results of an automatic cell-tracking algorithm\cite{Amat2014} are overlaid on the source data for visual inspection (\textbf{l}).}
  \label{fig:fig1}
\end{FPfigure}

\newpage
$ $
\vspace{-4mm}

\noindent
{\LARGE\textbf{Supplementary Videos}}
\vspace{1cm}

\noindent
\textbf{Supplementary Video 1.}\\[1mm]
\href{http://fly.mpi-cbg.de/~pietzsch/bdv-videos/supplementary-video-1.mp4}{\url{http://fly.mpi-cbg.de/~pietzsch/bdv-videos/supplementary-video-1.mp4}}\\[1mm]
Basic \bdv functionality, demonstrated on a large local XML/HDF5 dataset.
The dataset is a $500\text{GB}$ SPIM time-lapse of \emph{drosophila melanogaster} embryogenesis, comprising $250$ timepoints with $6$ angles each.
The data was acquired on a Zeiss Lightsheet Z.1 microscope.
\vspace{1cm}

\noindent
\textbf{Supplementary Video 2.}\\[1mm]
\href{http://fly.mpi-cbg.de/~pietzsch/bdv-videos/supplementary-video-2.mp4}{\url{http://fly.mpi-cbg.de/~pietzsch/bdv-videos/supplementary-video-2.mp4}}\\[1mm]
Demonstrates access to online CATMAID image datasets.
\vspace{1cm}

\noindent
\textbf{Supplementary Video 3.}\\[1mm]
\href{http://fly.mpi-cbg.de/~pietzsch/bdv-videos/supplementary-video-3.mp4}{\url{http://fly.mpi-cbg.de/~pietzsch/bdv-videos/supplementary-video-3.mp4}}\\[1mm]
Visualizing one timepoint of a multi-angle SPIM recording, before registration, after registration, and after deconvolution~\cite{sup:deconv} with Fiji's SPIM image processing tools.
\vspace{1cm}

\newpage
\renewcommand{\contentsname}{Supplementary Notes}
\setcounter{tocdepth}{1}
\tableofcontents

\chapter{Rendering}

\section{Introduction}
\Bdv is a re-slicing browser.
It takes a set of image volumes that are registered into a common global space and displays an arbitrary slice through that global space.
This supplement explains the details of this rendering procedure.
Several steps are necessary for this.
The source image volumes need to be transformed into a global coordinate system.
For each rendered pixel on the screen, the source voxels that contribute to it need to be determined.
Voxel intensities need to be converted from the space they are defined in to RGB color space for display on the screen,
  and colors contributed by different source volumes must be blended to a final output color.

To perform these operations we rely heavily on ImgLib2~\cite{sup:imglib2}, a generic image processing library which we use both
  to represent image volumes and implement slice rendering.
Section~\ref{sec:render:imglib2} reviews important ImgLib2 features and discusses a basic rendering algorithm.
With data that does not fit into the memory of the rendering computer, we need to take data caching, loading bandwith and latency into consideration.
In Section~\ref{sec:rendernonblocking} we discuss how we handle this in a non-blocking data loading scheme.
\Bdv uses multi-resolution image pyramids as one strategy to facilitate interactive browsing in the face of extremely large datasets and bandwith limitations.
In Section~\ref{sec:rendermipmap} we devise a refined rendering algorithm that leverages multi-resolution volumes and non-blocking data loading,
  and is extendable to custom, user-defined data sources displaying annotations or processing results.

\newcommand\tfsup[2]{\ensuremath{\mat{#1}^{\mathsf{#2}}}\xspace}
\newcommand\vcsup[2]{\ensuremath{\vc{#1}^{\mathsf{#2}}}\xspace}
\newcommand\cfrm[1]{\ensuremath{\mathsf{#1}}\xspace}

\newcommand\Tlm{\tfsup{T}{LM}}
\newcommand\Tlms[1]{\tfsup{T}{LM_{#1}}}
\newcommand\Twl{\tfsup{T}{WL}}
\newcommand\Tvw{\tfsup{T}{VW}}
\newcommand\Tproj{\ensuremath{\mat{P}}\xspace}

\newcommand\xl{\vcsup{x}{L}}
\newcommand\xw{\vcsup{x}{W}}
\newcommand\xv{\vcsup{x}{V}}
\newcommand\xm{\vcsup{x}{M}}
\newcommand\xs{\vcsup{x}{S}}

\section{Rendering using ImgLib2}
\label{sec:render:imglib2}

\subsection{ImgLib2}
In \bdv, we employ the generic image processing library ImgLib2~\cite{sup:imglib2} to represent image volumes and implement the slice rendering algorithm.
ImgLib2 allows to express algorithms in a way that abstracts from the data type, dimensionality, or memory storage of the image data.
For \bdv we rely in particular on the following key features:
\begin{itemize}
\item virtualized pixel access,
\item transparent, virtualized image extension,
\item transparent, virtualized image interpolation,
\item transparent, virtualized coordinate transformations.
\end{itemize}

The first, virtualized pixel access, provides the basis for the latter three.
Virtualized access means that in ImgLib2 the pixels of an image are accessed through an abstraction layer that hides storage details such as the memory layout used to store pixel values.
This allows images to be backed by arbitrary storage mechanisms.
One storage scheme provided by ImgLib2 is the |CellImg|, an image container that stores an $n$-dimensional image as a set of flat arrays, each representing a ($n$-dimensional hyper-)block of the image.
In this example, virtualized access hides the logic of which pixel is stored in which flat array, in a way that is completely transparent to algorithms accessing the pixels.

In \bdv, we build on the |CellImg| storage scheme and extend it to provide cache-backed image volumes.
Again, individual blocks are stored as flat arrays.
However, not all blocks are in memory all the time.
Instead, blocks are loaded on demand and cached using in a most-recently-used scheme.
The load-on-demand is triggered by virtualized pixel access: If an algorithm accesses a pixel that is not currently in memory then the corresponding block is loaded and cached.
This is extremely convenient for our rendering algorithm that can operate under the assumption that all data is in memory all the time.

Virtualized access provides the basis for virtualized image transformations.
Instead of being stored as pixel arrays in memory, images can be backed by transformation into other images.
This allows transparent image transformations that are lazily evaluated.
Only when a pixel is accessed, its coordinates are transformed into the original image space and the corresponding original pixel is accessed.
This allows for the latter three features listed above.

Virtualized image extension means that an image is extended to infinity by defining how values beyond the image boundaries are computed.
For example these may be fixed to a constant value, obtained by repeating or mirroring the image content, \etc.
Virtualized access takes care of generating out-of-bounds pixels using the specified rule.
For our rendering algorithm, we extend the raw image volumes with the background color.
The rendering algorithm does not have to consider whether it accesses pixels within or beyond image boundaries.

Virtualized image interpolation makes a discrete pixel image accessible at arbitrary real-valued coordinates.
ImgLib2 provides several interpolation schemes that define how values at non-integer coordinates are interpolated.
For \bdv we currently use nearest-neighbor an trilinear interpolation.

Virtualized coordinate transformations are used to transform an image into another coordinate frame.
The transformed image is transparent, \ie, coordinate transformation are performed on-the-fly when accessing pixels.
No data is copied, accessing a pixel of the transformed image means accessing the corresponding pixel of the original image.
In \bdv, we use this to spatially calibrate non-isotropic microscopy acquisitions, to register raw image volumes into a common global coordinate system, and to map the global coordinate system into the desired virtual slice for rendering, as detailed in the next section.

What makes the facilities described above even more powerful is that they can be effortlessly combined and layered.
An image extended to infinity is again an image which can be interpolated or coordinate-transformed to yield yet another image, \etc.
Whether the underlying data lives in our cache-backed |CellImg| or in a standard memory array is irrelevant.

\subsection{Rendering Arbitrary Slices of a Multi-View Dataset}
\label{sec:renderbasic}

This section explains the basic algorithm for rendering a slice through a registered multi-view dataset.
Note that for timelapse datasets we only need to consider a single timepoint.
For the moment, we assume that there is a single image volume corresponding to each view.
Handling multi-resolution data, where each image volume is available in progressively down-scaled resolutions (mipmaps), will be discussed in Section~\ref{sec:rendermipmap}.

We will represent coordinates in three coordinate frames:
\begin{itemize}
\item
The local frame \cfrm{L} of a raw image volume is defined by the 3D voxel coordinates of the volume.
\item
The global reference frame \cfrm{W} is an arbitrarily defined isotropic 3D coordinate system.
We denote by \Twl the transformation of local coordinates \xl to global coordinates \xw, that is, the registration of an image volume into the global reference frame.
\item
The viewer frame \cfrm{V} is a 3D coordinate system defined such that its $z=0$ plane coincides with the rendering canvas on the screen.
That is, the value at $(5,7,0)$ in the viewer frame will is rendered to pixel $(5,7)$ on the canvas.
We denote by \Tvw the transformation of global coordinates \xw to viewer coordinates \xv.
The transformation \Tvw represents the current rendering transform.
It is modified by the user by zooming, panning, reslicing, \etc,
\end{itemize}

\begin{figure}
\centerline{\includegraphics{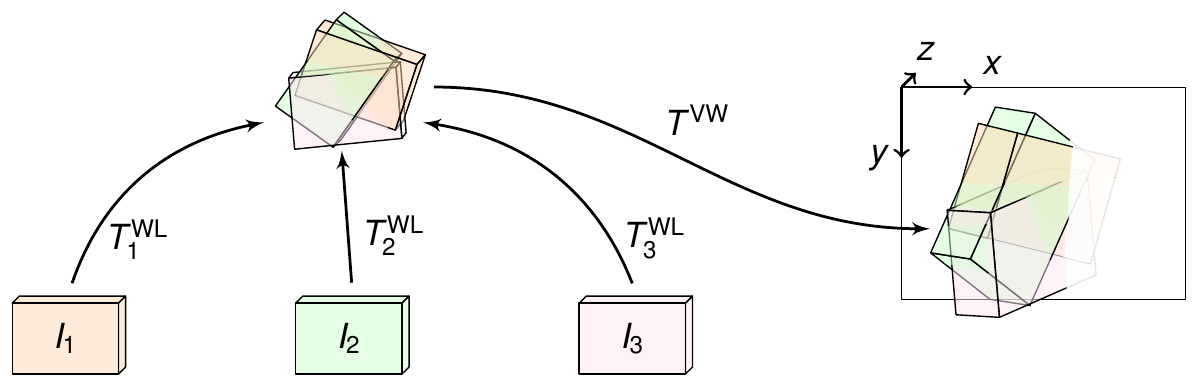}}
\caption{Coordinate transformations involved in rendering a slice through a set of registered volumes.
  Volumes $I_i$ are transformed into a common global reference frame using local-to-global transformations $\Twl_i$.
  Then the viewer transformation \Tvw is applied to transform global coordinates into the viewer frame.
  The plane $z=0$ of the viewer frame coincides with the rendering canvas on the screen.
  Pixels are rendered by copying this plane to the rendering canvas.}
\label{fig:render:transforms}
\end{figure}

Assume that the following inputs are given:
\begin{itemize}
  \item
    A rendering transform \Tvw, mapping global to viewer coordinates.
  \item
    $n$ raw image volumes $I_1, \dots, I_n$ of pixel types $\tau_1, \dots, \tau_n$.
    (For example, image data from \bdv HDF5 files has pixel type \emph{unsigned short}.)
  \item
    $n$ transformations $\Twl_1, \dots, \Twl_n$, where $\Twl_i$ maps voxel coordinates in $I_i$ to global coordinates.
  \item
    $n$ converters $C_1, \dots, C_n$, where a \emph{converter} $C_i$ is a function that maps values of pixel type $\tau_i$ to RGB pixel values for display on screen.
  \item
    A 2D RGB canvas $O$ for rendering pixels to the screen.
\end{itemize}
The basic algorithm for rendering to the canvas can be now formulated:
Every raw image volume is virtually extended and interpolated to be continuously accessible, \ie, pixel values are defined for every real-valued coordinate.
Then we (virtually) transform coordinates, first to the global reference frame using \Twl and then to the viewer frame using \Tvw.
Then we (virtually) convert pixel values from the pixel type of the raw volume to pixel type RGB for display.
Now we can simply read pixels of slice $z=0$ of the transformed images, combine them using a blending function $\mathcal{B}$, and write to the canvas.  Currently, we use a simple summation ($\sum$) as the blending function.

More formally, the basic rendering procedure is expressed in Algorithm~\ref{alg:render}.
The coordinate transformation steps used by the rendering algorithm are illustrated in Figure~\ref{fig:render:transforms}.

\begin{algorithm}[h]
  \DontPrintSemicolon
  \SetKwInOut{Input}{Input}\SetKwInOut{Result}{Result}
  \Input{viewer transform $\Tvw$,\\
    source volumes $I_1, \dots, I_n$,\\
    source transforms $\Twl_1, \dots, \Twl_n$,\\
    source converters $C_1, \dots, C_n$,\\
    canvas $O$}
  \Result{rendered image in $O$}
  \BlankLine
  \For{$i \in \{1, \dots, n\}$}{
    $I_i' := \operatorname{interpolate} \left( \operatorname{extend} \left( I_i \right) \right)$\;
    $I_i'' := \Tvw \left( \Twl \left( I_i' \right) \right)$\;
    $J_i := C_i \left( I_i'' \right)$\;
  }
  \For{{\bf every canvas pixel} $(x,y) \in O$}{
    set $O(x,y) := \mathcal{B}_{i=1}^{n} J_i \left( x, y, 0 \right)$\;
  }
  \caption{Basic rendering algorithm.}\label{alg:render}
\end{algorithm}

\section{Non-Blocking Data Access}
\label{sec:rendernonblocking}
Representing raw image volumes as cache-backed images is necessitated by the fact that we need to render datasets that
  are far too large to fit into the available memory of the rendering computer.
We will often face the situation that a voxel needed for rendering is not currently in cache and needs to be
  loaded from a data source, \eg, an HDF5 file on the local disk or a on-line data store.
In this case, we have to decide how to handle a request for such a voxel.
The simplest solution would be to use blocking access: When an uncached voxel is requested, a loading operation is
  triggered and the access blocks until the voxel data is available.
However, with on-line data sources in particular, this approach is problematic.
When data has to be fetched over unreliable network connections we cannot make guarantees regarding the bandwith
  or latency of the loading operation.

Using blocking access, a rendering operation might be blocked indefinitely.
For interactive rendering it is not desirable to block while waiting for data.
For the purpose of immediate interactive feedback it is very much preferable to present the user with partial data
  instead of rendering the application unresponsive while waiting for complete data.
To handle this requirement and to provide an alternative to blocking access, we introduce |Volatile| pixel types into ImgLib2~\cite{sup:imglib2}.

|Volatile| pixel types represent each pixel value as a tuple.
The tuple comprises an intensity value for the pixel, and a validity flag that signals whether the intensity value is valid (\ie, in our case
  whether it exists in memory or is still waiting to be loaded).
This allows to implement a deferred loading scheme that provides immediate feedback.

Let us consider a concrete example.
Image data is stored in \bdv HDF5 files with 16-bit precision.
Assuming a blocking scheme, this means that our cache-backed image volumes have |ShortType| pixel values in ImgLib2 terms.
With the deferred loading scheme, the cache-backed image instead has pixel type |VolatileShortType|.
When a voxel is accessed that is not currently in cache, the storage memory for the voxel's block is immediately allocated
  and the voxel is immediately ready to be processed.
However, the validity flag associated with the voxel is |false|, indicating that the voxel's intensity data is not (yet) valid.
The loading of intensity data is carried out asynchronously in a background thread.
Once the data is loaded, the validity flag changes to |true|, indicating that the intensity data is now valid.

Rendering from such |Volatile| cache-backed images is always possible without delay, presenting the user with partial data and immediate feedback.
In most cases, the majority of the required data will be available in cache already.

Crucially, arithmetic operations on |Volatile| types are implemented to correctly propagate validity information.
For example, an interpolated value computed from several voxel values will only be valid if all participating voxel values were valid.
This enables the rendering algorithm to track the number of invalid pixels that were rendered to the screen,
  and repeatedly render the image until all pixels are valid.

\section{Rendering Multi-Resolution Sources}
\label{sec:rendermipmap}
In \bdv, raw image volumes are typically available as multi-resolution pyramids.
In addition to the original resolution there exist several progressively down-scaled versions (mipmaps).
There are two main reasons for using multiple mipmap levels.
First, aliasing effects in zoomed-out views can be reduced by choosing the appropriate mipmap level to render.
Second, low-resolution versions occupy less memory and can therefore be transferred faster from disk or over a network connection.

\Bdv makes use of this by rapidly loading and rendering low-resolution data to provide immediate feedback when the user browses portions of the dataset that are not yet in cache.
Higher-resolution details are filled in as time permits, until the data is rendered at the optimal resolution level.
Section~\ref{sec:renderoptimipmap} discusses what we mean by ``optimal'' and how the optimal mipmap level is determined.

To make best use of the cache, we allow resolution levels to stand in for each other.
When trying to render voxels from a particular mipmap level, parts of the data that are missing in this particular mipmap level are replaced with data that is present in other mipmap levels.
Section~\ref{sec:rendermipmapalgo} describes how this is implemented in the rendering algorithm.

\subsection{Choosing the Optimal Mipmap Level}
\label{sec:renderoptimipmap}
Given a multi-resolution pyramid of an image volume, we want to select the resolution level for rendering that will provide the best quality.
Intuitively, we should choose the level that is closest to the on-screen resolution of the rendered image.
More formally, the ratio between the source voxel size (projected to the screen) and the screen pixel size should be close to 1.
Unfortunately, ``projected source voxel size'' is not well-defined.
The source resolution may be anisotropic and therefore the width, height, and depth of a voxel projected to the screen may yield different ``sizes''
In the following, projected source voxel size refers to the largest of these values.

\newcommand\Tsm{\ensuremath{\mat{T}}\xspace}

Let us assume that image volume $I$ is available in mipmap levels of different resolutions.
Let $\Tlms{k}$ denote the transformation from voxel coordinates in the $k^\text{th}$ mipmap level to voxel coordinates in the (full-resolution) image volume.
Let $\Tproj(\cdot)$ denote 3D-to-2D projection.
Then voxel coordinates $\xm$ in the $k^\text{th}$ mipmap level transform to screen coordinates \xs as
\begin{displaymath}
  \xs = \Tproj \left( \Tvw \left( \Twl \left( \Tlms{k} \left( \xm \right) \right) \right) \right) = \Tsm_k \left( \xm \right)
\end{displaymath}
where we use $\Tsm_k$ to denote the concatenated chain of transformations.
Let $\vc{o}=(0,0,0)$, $\vc{i}=(1,0,0)$, $\vc{j}=(0,1,0)$, $\vc{k}=(0,0,1)$ denote the origin and unit vectors along the $X$, $Y$, $Z$ axes, respectively.
Then the projected source voxel size is
\begin{displaymath}
  s_k = \max_{\vc{u} \in \{\vc{i}, \vc{j}, \vc{k} \}} \| \Tsm_k \left( \vc{u} \right)  -  \Tsm_k \left( \vc{o} \right) \|.
\end{displaymath}
The projected source voxel size can be used to select a single mipmap level for rendering by choosing $k$ that minimizes \UndefineShortVerb{\|}$|1-s_k|$\DefineShortVerb{\|}.

\subsection{Rendering with Mipmapped Volatile Sources}
\label{sec:rendermipmapalgo}
Rendering from unreliable multi-scale data sources with non-blocking access requires and alternative strategy to selecting the single best resolution as that single best resolution may not be available for indefinite time while others could be used temporarily.  We have implemented the following strategy to cope with this situation: All available mipmap levels are sorted by their expected rendering quality.  The order of this sorted list depends on both zoom and pose of the current virtual slice and is therefore updated at each change in pose or zoom.
Assume that raw image volumes are available as multi-resolution pyramids, where each mipmap level is a cache-backed image of |Volatile| pixel type.
Remember that this means that each voxel of each mipmap level has a validity flag that indicates whether the voxel's intensity value is currently in memory or whether the voxel is still pending to be loaded.
When rendering a pixel on screen, we go through the list of mipmap levels starting from the best entry and try to compute a value for the rendered pixel.
If the pixel cannot be rendered with valid data from the best mipmap level, we can try to render using the second best mipmap level and so on.

The resulting rendering algorithm is specified in Algorithm~\ref{alg:rendermain} (which uses the auxiliary \emph{RenderView} procedure specified in Algorithm~\ref{alg:renderview}).
We use the notation introduced in Sections~\ref{sec:renderbasic} and~\ref{sec:renderoptimipmap}, with the following augmentation.
Arguments $I_1, \dots, I_n$ to Algorithm~\ref{alg:rendermain} denote \emph{ordered mipmap pyramids}.
That is, $I_i$ is a list of mipmap levels for view $i$ that should be considered for rendering, ordered by quality from best to worst.
We assume that the list contains $m$ mipmap levels.
We denote by  $I_i^k$ the $k^\text{th}$ mipmap level, \ie, a (possibly down-scaled) image volume.
We denote by $\Tlms{k}_i$ the transformation from voxel coordinates in the $k^\text{th}$ mipmap level to voxel coordinates in the (full-resolution) image volume.

\begin{algorithm}[h]
  \DontPrintSemicolon
  \SetKwInOut{Input}{Input}
  \SetKwInOut{Result}{Result}
  \SetKwProg{Procedure}{Procedure}{:}{end}
  \Procedure{RenderView($\;\Tvw, \Twl, I^1, \dots, I^m, \Tlms{1}, \dots, \Tlms{m}, C, O, V$)}{
  \BlankLine
  \Input{viewer transform $\Tvw$,\\
    source transform $\Twl$\\
    mipmap levels $I^1, \dots, I^m$,\\
    mipmap transforms $\Tlms{1}, \dots, \Tlms{m}$,\\
    converter $C$,\\
    canvas $O$,\\
    validity mask $V$}
  \Result{partially rendered image in $O$,\\ updated validity mask in $V$}
  \BlankLine
  \For{$k \in \{1, \dots, m\}$}{
    $I' := \operatorname{interpolate} \left( \operatorname{extend} \left( I^k \right) \right)$\;
    $I'' := \Tvw \left( \Twl \left( \Tlms{k} \left( I' \right) \right) \right)$\;
    $J := C \left( I'' \right)$\;
    \For{{\bf every canvas pixel} $(x,y) \in O$}{
      \If{$V \left( x, y \right) \geq k$}{
        \If{$J^i \left( x, y, 0 \right)$ {\bf is valid}}{
          $O(x,y) := J \left( x, y, 0 \right)$\;
          $V(x,y) := k - 1$\;
        }
      }
    }
  }
  }
  \caption{\emph{RenderView} procedure. This partially renders one view (source multi-resolution pyramid).
    Available data from all mipmap levels is used.
    For each rendered pixel $(x,y)$, the next-best mipmap level which could be used to improve the pixel is written to $V(x,y)$.
    That is, $V(x,y)=0$ implies that the pixel $(x,y)$ has been rendered with the best possible quality.}\label{alg:renderview}
\end{algorithm}


\begin{algorithm}[h]
  \DontPrintSemicolon
  \SetKwInOut{Input}{Input}
  \SetKwInOut{Result}{Result}
  \Input{viewer transform $\Tvw$,\\
    ordered source pyramids $I_1, \dots, I_n$,\\
    source transforms $\Twl_1, \dots, \Twl_n$,\\
    source converters $C_1, \dots, C_n$,\\
    canvas $O$}
  \Result{rendered image in $O$}
  \BlankLine
  \tcp{initialization}
  \For{$i \in \{1, \dots, n\}$}{
    create empty canvas $O_i$\;
    create empty mask image $V_i$\;
    set $V_i(x,y) := m_i$ for all $(x,y) \in O$\;
  }
  \BlankLine
  \tcp{main render-and-display loop}
  \Repeat{$\sum_i \sum_{(x,y) \in O} V_i(x,y) = 0$}{
    \BlankLine
    \For{$i \in \{1, \dots, n\}$}{
      \If{$\sum_{(x,y) \in O} V_i(x,y) > 0$}
      {
        \emph{RenderView($\Tvw, \Twl_i, I_i^1, \dots, I_i^{m_i}, \Tlms{1}_i, \dots, \Tlms{m_i}_i, C_i, O_i, V_i$)}\;
      }
    }
    \BlankLine
    \For{{\bf every canvas pixel} $(x,y) \in O$}{
      set $O(x,y) := \mathcal{B}_{i=1}^{n} O_i(x,y)$\;
    }
    \BlankLine
    \emph{display($O$)}\;
    \BlankLine
  }
  \caption{Rendering algorithm.
      In the main loop we partially render all views and sum the rendered views into a partially rendered final image for display.
      This is repeated until all pixels are rendered at the desired optimal resolution level.
      Note that the loop may be prematurely aborted with an incomplete image if the user navigates away from the current slice.
  }\label{alg:rendermain}
\end{algorithm}

Verbally, the algorithm can be summarized as follows.
We first consider a single view (\ie, the ordered mipmap pyramid for one source image volume).
We create a mask image of the same size as the render canvas.

The mask image contains for each rendered pixel the largest index $k$ of a mipmap level that could be used to improve it (if its data were valid).
All mask pixels are initialized to $m$, meaning that the pixel is not rendered at all and could therefore be improved by any mipmap level.

For each mipmap level $k$ (starting from the best) we go over the render canvas.
For every pixel we check whether it was already drawn with the current mipmap level (or a better one).
If not, we check whether the current mipmap level has valid data for the pixel (this will either result in valid data or trigger asynchronous loading of the data).
If the data is valid, we set the pixel in the output image and set the corresponding mask pixel to $k-1$.
Then we go to the next mipmap level until all pixels have been drawn once, or there is no more mipmap level.

To overlay multiple views, the above procedure is used for each view to render an intermediate result.
The results are blended into the final rendered image.
For each of the single-view results we check whether all pixels have been drawn at the optimal mipmap level, \ie, $k=0$.
If not the whole procedure is repeated.
Note, that already-completed views need not be rendered again in repeat passes.

We employ two additional tweaks to reduce rendering artifacts that have been omitted from the above discussion.
These require a tighter integration between the rendering code and the cache.
Without these tweaks the following artifact may occur.
Assume, that in the first render pass there are voxels that are missing from the cache \emph{for all mipmap levels}.
Then the rendering algorithm will trigger asynchronous loading of the missing data by touching these voxels.
For a short period, until the data is loaded, these voxels will remain invalid in all mipmap levels.
This will result in pixels that cannot be rendered at all and therefore will appear as black artifacts on the screen.

To remedy this, we do the following.
First, before starting to render, we do a fast prediction of which data chunks will be touched by the rendering.
We trigger loading of these chunks such that chunks from the lowest-resolution mipmap will be loaded first.
The intuition for this is that for the lowest-resolution mipmap, we will need to load only little data, which will happen very fast.
Therefore, when the rendering starts it is likely that the low-resolution data is in cache and we will be able to successfully render (at least a low-resolution version of) each pixel.
Second, we reserve a small time budget during which loading operations are allowed to block.
That is, until the time budget is used up, when an invalid voxel is hit, we allow for a small delay during which the voxel might become valid.

\section{Interactive Navigation at High Resolution}
\label{sec:interactive}

On a modern display, a virtual slice at full resolution can easily comprise several millions of pixels.  While \bdv's slice rendering is very efficient, on current hardware, the update rate at full resolution is not always satisfying.  In order to achieve truly interactive browsing experience, \bdv renders lower resolution slices first and gradually improves the resolution as the slice stops moving.  \bdv's renderer permanently measures the time required to generate a slice at each given resolution and sets the lowest resolution for interactive browsing to the maximum resolution at which update rates of $> 20$ frames per second can be guaranteed.  This way, we combine the best of two worlds, interactive smooth navigation and high quality still images for detailed inspection.

\section{Extensibility}
\label{sec:renderextensibility}
\Bdv's rendering algorithm is designed with our caching infrastructure in mind.
It is aware of multi-resolution pyramids, |Volatile| voxel types, and cache-backed |CellImg| volumes.
%
However, it is important to note that caching and rendering in \bdv are only loosely coupled.
In fact, the renderer will take advantage of data sources that are backed by a cache, provide multiple resolutions, or have |Volatile| voxel type.
But data sources do not need to have these properties.

We took care to make it easy to add additional data sources to the renderer, for example to overlay segmentations and similar processing results.
Any image that exposes a standard ImgLib2 interface can be trivially wrapped as a data source.
The rendering algorithm of course works with non-mipmapped sources, because these can be treated as resolution pyramids with only a single level.
Sources that do not expose |Volatile| voxel types are also trivially handled, because their voxels are \emph{always} valid.
Finally, sources that are infinitely large or not restricted to an integer grid can be rendered just as well.
For these we can simply omit interpolation and extension that are required for the standard bounded, rasterized sources.

\begin{figure}
\centerline{\includegraphics[width=.8\textwidth]{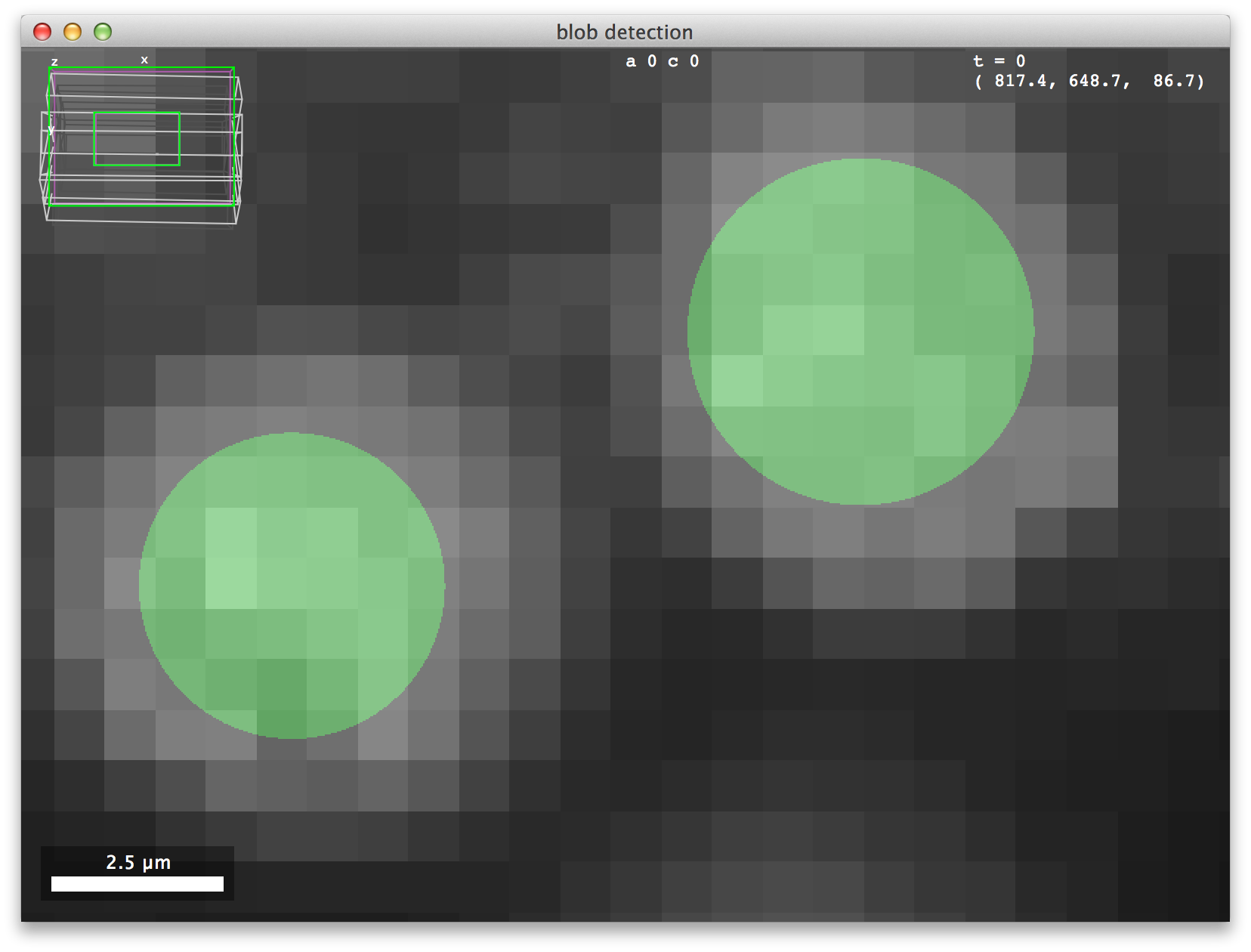}}
\caption{Rendering of custom sources.
  Visualisation blob-detection algorithm is added as an additional source for rendering in \bdv.
  The source is continuous and defined with a \texttt{boolean} pixel type:
  If a coordinate lies within a given radius of the center of a detected blob the associated value is \emph{true}, otherwise it is \emph{false}.
  A custom converter to RGB converts \emph{true} to green and \emph{false} to black.
  The zoomed-view illustrates that the source is continuous (\ie, has infinite resolution).
}
\label{fig:render:continuous}
\end{figure}

The ability to add custom sources is illustrated in Figure~\ref{fig:render:continuous}.
Here we render an additional custom source to visualize the results of a blob detection algorithm.
The custom source shows a virtualized image that is backed by a list of blob centers and radii.
This image is continuous, infinite, and defined with a \texttt{boolean} voxel type.
When a voxel is accessed, its value is determined on-the-fly by comparing its coordinate to the blob list.
If the coordinate lies within the blob radius of the center of a blob then the voxel value is \emph{true}, otherwise it is \emph{false}.
Note, that this image is unbounded and continuously defined.
The blob containment check can be performed for any real-valued coordinate, \ie, the image has infinite resolution.
To display the source, we provide a boolean-to-RGB converter that converts \emph{true} to green and \emph{false} to black.

\chapter{File Format}

\section{Introduction}

BigDataViewer provides a custom file-format that is optimized for fast arbitrary re-slicing at various scales.
The file format is build on open standards XML\cite{sup:xml} and HDF5\cite{sup:hdf5}, where HDF5 is used to store image volumes and XML is used to store meta-data.
Section~\ref{sec:ffoverview} gives a high-level overview of the file-format, Sections~\ref{sec:ffxml} and~\ref{sec:ffhdf5} provide more detail on the XML and HDF5 parts respectively.

The format is extensible in multiple ways:
The XML file of a dataset can be augmented with arbitrary additional meta-data. Fiji's SPIMage processing pipeline makes use of this for example, to store information about detected beads and nuclei.
Moreover, the HDF5 file of the dataset can be replaced by other storage methods, for example TIFF files or remote access to data available online.
Extensibility is further discussed in Section~\ref{sec:ffext}.

\section{Overview}
\label{sec:ffoverview}

Each \bdv dataset contains a set of 3D grayscale image volumes organized by \timepoints and \setups.
In the context of lightsheet microscopy, each channel or acquisition angle or combination of both is a \setup.
\Eg, for a multi-view recording with 3 angles and 2 channels there are 6 \setups.
Each \setup represents a visualisation data source in the viewer that provides one image volume per \timepoint.
We refer to each combination of \setup and \timepoint as a \view.
Each \view has one corresponding grayscale image volume.

A dataset comprises an XML file to store meta-data and one or more HDF5 files to store the raw images.
Among other things, the XML file contains
\begin{itemize}
  \item the path of the HDF5 file(s),
  \item a number of \emph{setups},
  \item a number of \emph{timepoints},
  \item the registration of each \emph{view} into the global coordinate system.
\end{itemize}

\begin{figure}
\centerline{\includegraphics[scale=0.9]{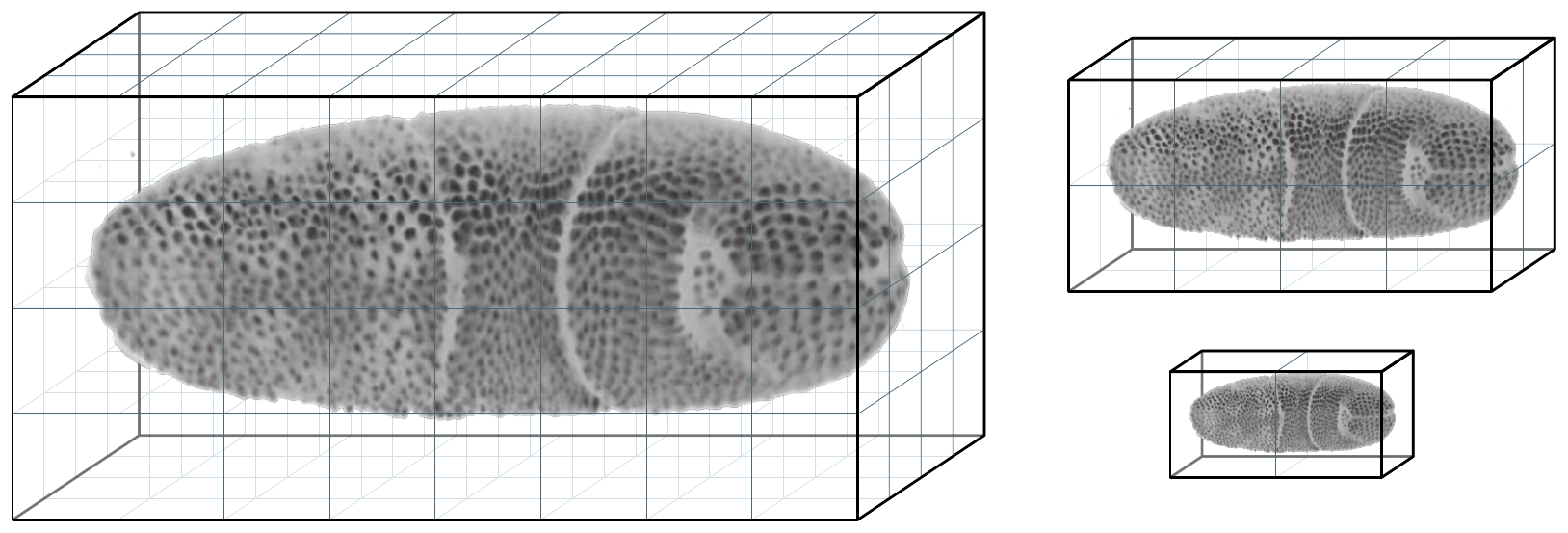}}
\caption{Chunked Mipmap Pyramid.
Each raw image volume is stored in multiple resolutions, the original resolution (left) and successively smaller, downsampled versions (right).
Each resolution is stored in a chunked representation, split into small 3D blocks.
}
\label{fig:multirespyramid}
\end{figure}

Each \view has one corresponding image volume which is stored in the HDF5 file.
Raw image volumes are stored as multi-resolution pyramids:
In addition to the original resolution, several progressively down-scaled resolutions (mipmaps) are stored.
This serves two purposes.
First, using mipmaps minimizes aliasing effects when rendering a zoomed-out view of the dataset~\cite{sup:Williams:1983it}.
Second, and more importantly, using mipmaps reduces data access time and thus increases the perceived responsiveness for navigation.
Low-resolution mipmaps take up less memory and therefore load faster from disk.
New chunks of data must be loaded when the user browses to a part of the dataset that is not currently cached in memory.
In this situation, \bdv can rapidly load and render low-resolution data, filling in high resolution detail later as it becomes available.
This multi-resolution pyramid scheme is illustrated in Figure~\ref{fig:multirespyramid}.

Each level of the multi-resolution pyramid is stored as a \emph{chunked multi-dimensional array}.
Multi-dimensional arrays are the standard way of storing image data in HDF5.
The layout of multi-dimensional arrays on disk can be configured.  We use a \emph{chunked} layout which means that the 3D image volume is split into several chunks (smaller 3D blocks).
These chunks are stored individually in the HDF5 file, which optimizes performance for our use-case where fast random access to individual chunks is required.

\begin{figure}
\centerline{\includegraphics{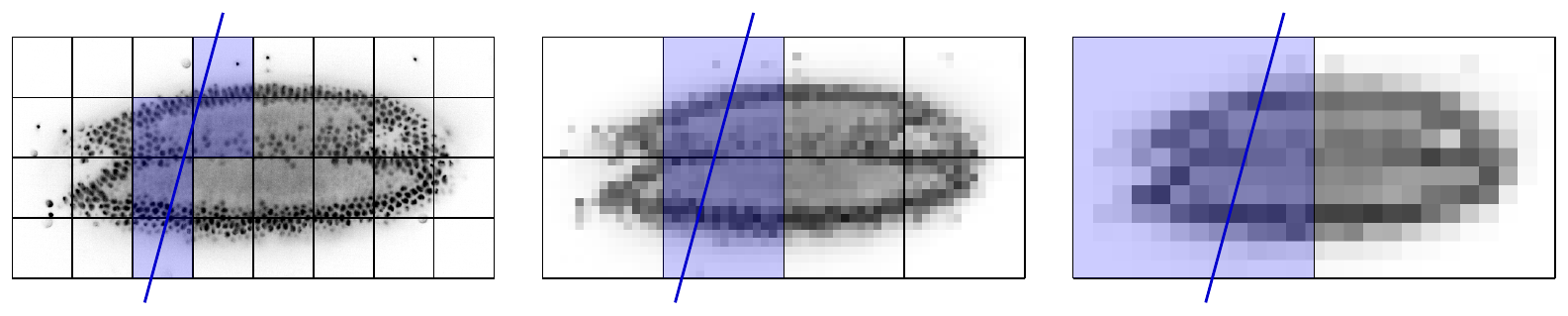}}
\caption{Loading Mipmap Chunks.
When rendering a slice (schematically illustrated by the blue line) the data of only a small subset of blocks is required.
In the original resolution 5 blocks are required, while only 2, respectively 1 block is required for lower resolutions.
Therefore, less data needs to be loaded to render a low-resolution slice.
This allows low-resolution versions to be loaded and rendered rapidly.
High-resolution detail is filled in when the user stops browsing to view a certain slice for an extended period of time.
}
\label{fig:blockloading}
\end{figure}

The performance of partial I/O, \ie. reading subsets of the data, is maximized when the data selected for I/O is contiguous on disk~\cite{sup:hdf5-chunk}.
The chunked layout is therefore well-suited to re-slicing access to images data.
Rendering a virtual slice requires data contained within a small subset of chunks.
Only chunks that touch the slice need to be loaded, see Figure~\ref{fig:blockloading}.
Each of these chunks, however, is loaded in full, although only a subset of voxels in each chunk is required to render the actual slice.  Loading the data in this way, aligned at chunk boundaries, gurantees optimal I/O performance.

\begin{figure}
\centerline{\includegraphics{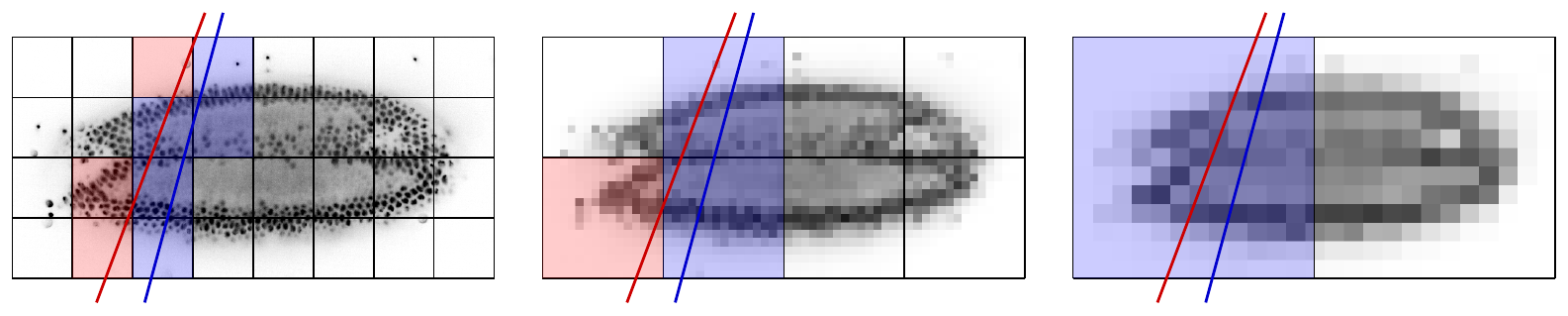}}
\caption{Caching Mipmap Chunks.
Recently used blocks are cached in RAM. For rendering the slice indicated by the red line, only the red blocks need to be loaded.
The blue blocks are already cached from rendering the blue slice before.
}
\label{fig:blockcaching}
\end{figure}

All loaded chunks are cached in RAM.
During interactive navigation, subsequent slices typically intersect with a similar set of chunks because their pose has changed only moderately, \ie. cached data are re-used.
Only chunks that are not currently in the cache need to be loaded from disk, see Figure~\ref{fig:blockcaching}.
Combined with the multi-resolution mipmap representation, this chunking and caching scheme allows for fluid interactive browsing of very large datasets.

The parameters of the mipmap and chunking scheme are specific to each dataset and they are fully configurable by the user.
In particular, when exporting images to the \bdv format, the following parameters are adjustable:
\begin{itemize}
  \item the number of mipmap levels,
  \item the subsampling factors in each dimension for each mipmap level,
  \item the chunk sizes in each dimension for each mipmap level.
\end{itemize}
\Bdv suggests sensible parameter settings, however, for particular applications and data properties a user may tweak these parameters for optimal performance.

\section{XML File Format}
\label{sec:ffxml}

In the following we describe the XML format by means of an example.
Consider a dataset that consists of 2 \setups and 3 \timepoints, that is, 6 \views in total.
The dataset can be specified in a minimal XML file as follows
\plset{language=xml}
\fvsetlist
\begin{Verbatim}[commandchars=\\\{\}]
\PY{c+cp}{\PYZlt{}?xml version=\PYZdq{}1.0\PYZdq{} encoding=\PYZdq{}UTF\PYZhy{}8\PYZdq{}?\PYZgt{}}
\PY{n+nt}{\PYZlt{}SpimData} \PY{n+na}{version=}\PY{l+s}{\PYZdq{}0.2\PYZdq{}}\PY{n+nt}{\PYZgt{}}
  \PY{n+nt}{\PYZlt{}BasePath} \PY{n+na}{type=}\PY{l+s}{\PYZdq{}relative\PYZdq{}}\PY{n+nt}{\PYZgt{}}.\PY{n+nt}{\PYZlt{}/BasePath\PYZgt{}}
  \PY{n+nt}{\PYZlt{}SequenceDescription\PYZgt{}}
    \PY{n+nt}{\PYZlt{}ImageLoader} \PY{n+na}{format=}\PY{l+s}{\PYZdq{}bdv.hdf5\PYZdq{}}\PY{n+nt}{\PYZgt{}}
      \PY{n+nt}{\PYZlt{}hdf5} \PY{n+na}{type=}\PY{l+s}{\PYZdq{}relative\PYZdq{}}\PY{n+nt}{\PYZgt{}}drosophila.h5\PY{n+nt}{\PYZlt{}/hdf5\PYZgt{}}
    \PY{n+nt}{\PYZlt{}/ImageLoader\PYZgt{}}
    \PY{n+nt}{\PYZlt{}ViewSetups\PYZgt{}}
      \PY{n+nt}{\PYZlt{}ViewSetup\PYZgt{}}
        \PY{n+nt}{\PYZlt{}id\PYZgt{}}0\PY{n+nt}{\PYZlt{}/id\PYZgt{}}
        \PY{n+nt}{\PYZlt{}name\PYZgt{}}angle 1\PY{n+nt}{\PYZlt{}/name\PYZgt{}}
      \PY{n+nt}{\PYZlt{}/ViewSetup\PYZgt{}}
      \PY{n+nt}{\PYZlt{}ViewSetup\PYZgt{}}
        \PY{n+nt}{\PYZlt{}id\PYZgt{}}1\PY{n+nt}{\PYZlt{}/id\PYZgt{}}
        \PY{n+nt}{\PYZlt{}name\PYZgt{}}angle 2\PY{n+nt}{\PYZlt{}/name\PYZgt{}}
      \PY{n+nt}{\PYZlt{}/ViewSetup\PYZgt{}}
    \PY{n+nt}{\PYZlt{}/ViewSetups\PYZgt{}}
    \PY{n+nt}{\PYZlt{}Timepoints} \PY{n+na}{type=}\PY{l+s}{\PYZdq{}range\PYZdq{}}\PY{n+nt}{\PYZgt{}}
      \PY{n+nt}{\PYZlt{}first\PYZgt{}}0\PY{n+nt}{\PYZlt{}/first\PYZgt{}}
      \PY{n+nt}{\PYZlt{}last\PYZgt{}}2\PY{n+nt}{\PYZlt{}/last\PYZgt{}}
    \PY{n+nt}{\PYZlt{}/Timepoints\PYZgt{}}
  \PY{n+nt}{\PYZlt{}/SequenceDescription\PYZgt{}}
  \PY{n+nt}{\PYZlt{}ViewRegistrations\PYZgt{}}
    \PY{n+nt}{\PYZlt{}ViewRegistration} \PY{n+na}{timepoint=}\PY{l+s}{\PYZdq{}0\PYZdq{}} \PY{n+na}{setup=}\PY{l+s}{\PYZdq{}0\PYZdq{}}\PY{n+nt}{\PYZgt{}}
      \PY{n+nt}{\PYZlt{}ViewTransform} \PY{n+na}{type=}\PY{l+s}{\PYZdq{}affine\PYZdq{}}\PY{n+nt}{\PYZgt{}}
        \PY{n+nt}{\PYZlt{}affine\PYZgt{}}0.996591 0.001479 0.010733 \PYZhy{}5.384684
        \PYZhy{}0.001931 0.995446 \PYZhy{}0.003766 \PYZhy{}81.544861
        \PYZhy{}0.000497 \PYZhy{}0.000060 3.490110 9.854919\PY{n+nt}{\PYZlt{}/affine\PYZgt{}}
      \PY{n+nt}{\PYZlt{}/ViewTransform\PYZgt{}}
    \PY{n+nt}{\PYZlt{}/ViewRegistration\PYZgt{}}
    \PY{n+nt}{\PYZlt{}ViewRegistration} \PY{n+na}{timepoint=}\PY{l+s}{\PYZdq{}0\PYZdq{}} \PY{n+na}{setup=}\PY{l+s}{\PYZdq{}1\PYZdq{}}\PY{n+nt}{\PYZgt{}}
\end{Verbatim}
\vspace{-1mm}
{\footnotesize\texttt{ ...}}
\fvset{firstnumber=66}
\begin{Verbatim}[commandchars=\\\{\}]
  \PY{n+nt}{\PYZlt{}/ViewRegistrations\PYZgt{}}
\PY{n+nt}{\PYZlt{}/SpimData\PYZgt{}}
\end{Verbatim}
\fvset{firstnumber=1}

\fvsettext
\SaveVerb{SpimData}|<SpimData>|
\SaveVerb{BasePath}|<BasePath>|
\SaveVerb{SequenceDescription}|<SequenceDescription>|
\SaveVerb{ImageLoader}|<ImageLoader>|
\SaveVerb{ViewSetups}|<ViewSetups>|
\SaveVerb{ViewSetup}|<ViewSetup>|
\SaveVerb{ViewRegistrations}|<ViewRegistrations>|
\SaveVerb{ViewRegistration}|<ViewRegistration>|
\SaveVerb{ViewTransform}|<ViewTransform>|
\SaveVerb{id}|<id>|
\SaveVerb{sname}|<name>|
\SaveVerb{Timepoints}|<Timepoints>|
\SaveVerb{ViewInterestPoints}|<ViewInterestPoints>|
\SaveVerb{Attributes}|<Attributes>|
\newcommand\SpimData{\UseVerb{SpimData}\xspace}
\newcommand\BasePath{\UseVerb{BasePath}\xspace}
\newcommand\SequenceDescription{\UseVerb{SequenceDescription}\xspace}
\newcommand\ImageLoader{\UseVerb{ImageLoader}\xspace}
\newcommand\ViewSetups{\UseVerb{ViewSetups}\xspace}
\newcommand\ViewSetup{\UseVerb{ViewSetup}\xspace}
\newcommand\ViewRegistrations{\UseVerb{ViewRegistrations}\xspace}
\newcommand\ViewRegistration{\UseVerb{ViewRegistration}\xspace}
\newcommand\ViewTransform{\UseVerb{ViewTransform}\xspace}
\newcommand\id{\UseVerb{id}\xspace}
\newcommand\sname{\UseVerb{sname}\xspace}
\newcommand\Timepoints{\UseVerb{Timepoints}\xspace}
\newcommand\ViewInterestPoints{\UseVerb{ViewInterestPoints}\xspace}
\newcommand\Attributes{\UseVerb{Attributes}\xspace}

The top-level \SpimData element contains at least a \BasePath element, a \SequenceDescription element, and a \ViewRegistrations element.
The ordering of elements is irrelevant.

\BasePath (line 3) defines the base path path for all relative paths occuring in the rest of the file.
Usually, this is ``|.|'', \ie, the directory of the XML file.

\SequenceDescription (lines 4--22) defines the \setups and \timepoints and thereby specifies the \views (raw image volumes) contained in the sequence.
It also specifies an \ImageLoader (line 5-7), which will be discussed later.
In the example we have two \ViewSetups (line 8--17).
Each \ViewSetup must have a unique \id (|0| and |1| in the example).
It may have a \sname (|angle 1| and |angle 2| in the example).
It may also have arbitrary additional attributes (see Section~\ref{sec:ffext}).
\Timepoints (lines 18--21) can be specified in several ways: as a range, as a list, or as a pattern.
In the example they are specified as a range, starting with \timepoint |0| and ending with \timepoint |2|.

\ViewRegistrations (lines 23--66) describe the transformations that register each \view's raw voxel coordinates into the global coordinate system.
In the example, there are 6 \views: (\timepoint 0, \setup 0) through (\timepoint 2, \setup 1).
Thus there are 6 \ViewRegistration child elements, one for each \view.
Each can be speficied as a single \ViewTransform 3d-affine matrix as in the example, or as a list of \ViewTransform elements which will be concatenated to obtain the final transform.

The \ImageLoader element (line 5-7)
describes a raw image volume source for each \view.
The default |<ImageLoader format="bdv.hdf5">| will read the volumes from an HDF5 file, as indicated by the |format| attribute.
The contents of the \ImageLoader element is specific to the format.
For HDF5, it simply specifies the name of the HDF5 file (line 6).

\section{HDF5 File Format}
\label{sec:ffhdf5}

\begin{figure}
\centerline{\includegraphics[width=.6\textwidth]{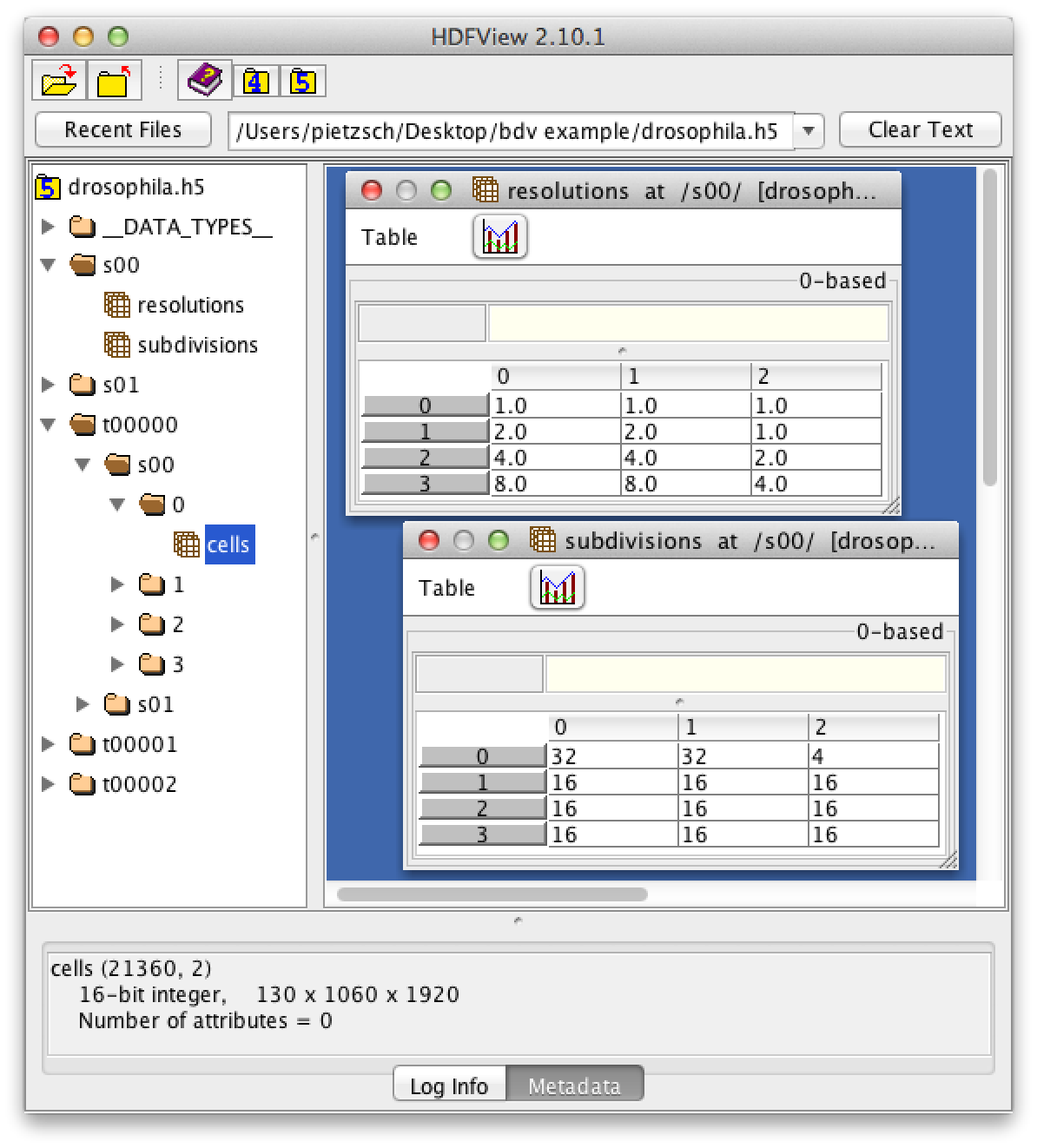}}
\caption{HDF5 File Structure. The HDF5 file of the example dataset shown in a HDF5 browser.
}
\label{fig:hdfview}
\end{figure}

The HDF5 file format is straightforward. It contains the chunked multi-dimen\-sional arrays for each \view of the dataset and a minimum of meta-data.
Figure~\ref{fig:hdfview} shows the HDF5 file of the example dataset, inspected in the standard \emph{HDFView} browser~\cite{sup:hdfview}.

The meta-data comprises the parameters of the mipmap pyramids.
We allow different parameters for each \setup, because the image volumes of individual \setups might be captured with different size, resolution, or anisotropy.
The parameters of a mipmap pyramid comprise the subsampling factors in each dimension for each mipmap level, and the chunk sizes in each dimension for each mipmap level.
The subsampling factors for the mipmap pyramid of the \setup with id \emph{SS} are stored as a matrix data object in the path ``\texttt{s}\emph{SS}\texttt{/resolutions}'' in the HDF5 file.
The chunk sizes of \setup \emph{SS} are stored as a matrix data object in the path ``\texttt{s}\emph{SS}\texttt{/subdivisions}''.
Having 2 \setups, the example file contains \texttt{s00/resolutions}, \texttt{s00/subdivisions}, \texttt{s01/resolutions}, and \texttt{s01/subdivisions}.
Consider \texttt{s00/resolutions} in the example dataset. This is the matrix
\begin{displaymath}
\begin{bmatrix}
  1 & 1 & 1 \\
  2 & 2 & 1 \\
  4 & 4 & 2 \\
  8 & 8 & 4
\end{bmatrix}
\end{displaymath}
where rows index the mipmap level and columns index the dimension.
For example, the 4$^\text{th}$ mipmap level has subsampling factors $8, 8, 4$ in dimension $X, Y, Z$ respectively.
Similary, \texttt{s00/subdivisions} is the matrix
\begin{displaymath}
\begin{bmatrix}
  32 & 32 & 4 \\
  16 & 16 & 16 \\
  16 & 16 & 16 \\
  16 & 16 & 16
\end{bmatrix}
\end{displaymath}
where rows index the mipmap level and columns index the dimension.
For example, the 0$^\text{th}$ mipmap level is chunked into blocks of size $32 \times 32 \times 4$ in $X \times Y \times Z$.

Image data is stored in exactly one chunked multi-dimensional array for every mipmap level for every \view.
These data arrays are stored in paths ``\texttt{t}\emph{TTTTT}\texttt{/s}\emph{SS}\texttt{/}\emph{L}\texttt{/cells}'' in the HDF5 file.
Here, \emph{TTTTT} is the id of the \timepoint, \emph{SS} is the id of the \setup, and \emph{L} is the index of the mipmap level.
Having 3 \timepoints, 2 \setups, and 4 mipmap levels, the example file contains \texttt{t00000/s00/0/cells} through \texttt{t00002/s01/3/cells}.
Currently, we always store image volumes with 16-bit precision.

\section{Extensibility}
\label{sec:ffext}

The \bdv XML format is extensible in several ways (which may seem self-evident because XML has ``eXtensible'' already in its name).
By extensible we mean the following:
The Java library that maps the XML file to a SpimData object representation in memory provides extension points for augmenting the XML (and object model) with additional content.\footnote{
  The SpimData Java library is open source and is available on \href{http://github.com/tpietzsch/spimdata}{\url{http://github.com/tpietzsch/spimdata}}.
}
Crucially, this is done in a backwards compatible way, such that different users of the format need not be aware of each others respective extensions.
For example, \bdv ignores parts of the XML file which are specific to Fiji's SPIMage processing tools.
Nevertheless it is able to read and write these files, leaving the SPIMage processing extensions intact.

In the following we briefly highlight the available XML extension points and their support by the SpimData Java library.

\subsection{Alternative Image Sources}
Instead of the default HDF5 backend, any other source providing image volumes may be specified.
As discussed in Section~\ref{sec:ffxml}, the type of image source is defined in the |format| attribute of the |<ImageLoader| |format="bdv.hdf5">| element.

\Bdv provides alternative backends in addition to HDF5, for example for accessing images provided by a CATMAID web service.
Adding a new type of image source requires
\begin{enumerate}
  \item a Java class |C| that implements the |BasicImgLoader<T>| interface (where |T| is the voxel type provided by the new format),
  \item a Java class that implements the |XmlIoBasicImgLoader<C>| interface and is annotated by |@ImgLoaderIo|, specifying the name of the new format.
\end{enumerate}
To give a concrete example, the implementation of the CATMAID web service backend consists of the classes
\plset{language=java}
\fvsetlist
\begin{Verbatim}[commandchars=\\\{\}]
\PY{k+kd}{public} \PY{k+kd}{class} \PY{n+nc}{CatmaidImageLoader} \PY{k+kd}{implements} \PY{n}{BasicImgLoader}\PY{o}{\PYZlt{}}\PY{n}{ARGBType}\PY{o}{\PYZgt{}}
\PY{o}{\PYZob{}} \PY{o}{...}
\PY{o}{\PYZcb{}}
\end{Verbatim}
\noindent
and
\begin{Verbatim}[commandchars=\\\{\}]
\PY{n+nd}{@ImgLoaderIo}\PY{o}{(}\PY{n}{format} \PY{o}{=} \PY{l+s}{\PYZdq{}catmaid\PYZdq{}}\PY{o}{,} \PY{n}{type} \PY{o}{=} \PY{n}{CatmaidImageLoader}\PY{o}{.}\PY{n+na}{class}\PY{o}{)}
\PY{k+kd}{public} \PY{k+kd}{class} \PY{n+nc}{XmlIoCatmaidImageLoader}
        \PY{k+kd}{implements} \PY{n}{XmlIoBasicImgLoader}\PY{o}{\PYZlt{}}\PY{n}{CatmaidImageLoader}\PY{o}{\PYZgt{}}
\PY{o}{\PYZob{}} \PY{o}{...}
\PY{o}{\PYZcb{}}
\end{Verbatim}
\fvsettext
\noindent
The actual implementations are beyond the scope of this document.\footnote{We refer to
  \href{http://github.com/tpietzsch/spimviewer}{\url{http://github.com/tpietzsch/spimviewer}}
  and
  \href{https://github.com/fiji/SPIM_Registration}{\url{https://github.com/fiji/SPIM_Registration}}
  which provide multiple example backends.
  Implementing a backend is particularly easy if the custom image format is able to load chunks of image volumes.
  We provide facilities that make it straightforward to re-use our caching infrastructure in this case.
}
All annotated |XmlIoBasicImgLoader| subclasses will be picked up automatically and used to instantiate |BasicImgLoader| when the specified format name is encountered.
For example, if an |<ImageLoader format="catmaid">| element is encountered it will be passed to |XmlIoCatmaidImageLoader|, which then will create a |CatmaidImageLoader|.

\subsection{Adding Custom SpimData Sections}
\label{sec:ffcustomsections}

Arbitrary top-level elements may be added to the \SpimData root element.
The only restriction is that each top-level element may occur only once.
As discussed in Section~\ref{sec:ffxml}, the elements \BasePath, \SequenceDescription, and \ViewRegistrations must always exist.
Fiji's SPIMage processing for example, adds a custom top-level element \ViewInterestPoints.
\plset{language=xml}
\fvsetlist
\begin{Verbatim}[commandchars=\\\{\}]
\PY{c+cp}{\PYZlt{}?xml version=\PYZdq{}1.0\PYZdq{} encoding=\PYZdq{}UTF\PYZhy{}8\PYZdq{}?\PYZgt{}}
\PY{n+nt}{\PYZlt{}SpimData} \PY{n+na}{version=}\PY{l+s}{\PYZdq{}0.2\PYZdq{}}\PY{n+nt}{\PYZgt{}}
  \PY{n+nt}{\PYZlt{}BasePath} \PY{n+na}{type=}\PY{l+s}{\PYZdq{}relative\PYZdq{}}\PY{n+nt}{\PYZgt{}}.\PY{n+nt}{\PYZlt{}/BasePath\PYZgt{}}
  \PY{n+nt}{\PYZlt{}SequenceDescription\PYZgt{}}
    ...
  \PY{n+nt}{\PYZlt{}/SequenceDescription\PYZgt{}}
  \PY{n+nt}{\PYZlt{}ViewRegistrations\PYZgt{}}
    ...
  \PY{n+nt}{\PYZlt{}/ViewRegistrations\PYZgt{}}
  \PY{n+nt}{\PYZlt{}ViewInterestPoints\PYZgt{}}
    ...
  \PY{n+nt}{\PYZlt{}/ViewInterestPoints\PYZgt{}}
\PY{n+nt}{\PYZlt{}/SpimData\PYZgt{}}
\end{Verbatim}
\fvsettext
\noindent
To be able to read and write files with this additional top-level element, the SPIMage processing tools use a custom reader/writer class.
All such reader/writer classes are derived from |XmlIoAbstractSpimData| which also takes care of unknown top-level elements.

A particular reader/writer may not be able to handle a particular top-level element.
For example the \bdv does not know how to handle the \ViewInterestPoints and therefore ignores it.
It would not be reasonable to expect every consumer of the XML format to understand additional content that is of no interest to them.
However, neither should additional content be simply discarded.
Otherwise it might get lost if a load-modify-save operation is performed on a file with additional content.

Instead, the |XmlIoAbstractSpimData| reader/writer stores unhandled top-level elements when opening a file.
If the same reader/writer is later used to write the (modified) dataset, this information is simply appended to the newly created file as-is.
This allows programmatic modification of datasets without understanding all extensions.
In summary, we avoid losing information while also avoiding the need for every consumer to handle every extension.

\subsection{Adding Custom \Setup Attributes}

The \ViewSetups section may be augmented with arbitrary \ViewSetup attributes to provide additional meta-data for the \setups.
While the \bdv requires no \setup attributes at all, Fiji's SPIMage processing requires at least the acquisition angle, channel, and illumination direction of the microscope.
Conceptually a particular attribute is a set of values.
These can be defined in \Attributes elements and particular attribute values may be associated to \setups using value ids.
To illustrate this, here is how the \emph{angle} attribute is defined and used.
\plset{language=xml}
\fvsetlist
\begin{Verbatim}[commandchars=\\\{\}]
\PY{c+cp}{\PYZlt{}?xml version=\PYZdq{}1.0\PYZdq{} encoding=\PYZdq{}UTF\PYZhy{}8\PYZdq{}?\PYZgt{}}
\PY{n+nt}{\PYZlt{}SpimData} \PY{n+na}{version=}\PY{l+s}{\PYZdq{}0.2\PYZdq{}}\PY{n+nt}{\PYZgt{}}
  \PY{n+nt}{\PYZlt{}BasePath} \PY{n+na}{type=}\PY{l+s}{\PYZdq{}relative\PYZdq{}}\PY{n+nt}{\PYZgt{}}.\PY{n+nt}{\PYZlt{}/BasePath\PYZgt{}}
  \PY{n+nt}{\PYZlt{}SequenceDescription\PYZgt{}}
    \PY{n+nt}{\PYZlt{}ViewSetups\PYZgt{}}
      \PY{n+nt}{\PYZlt{}ViewSetup\PYZgt{}}
        \PY{n+nt}{\PYZlt{}id\PYZgt{}}0\PY{n+nt}{\PYZlt{}/id\PYZgt{}}
        \PY{n+nt}{\PYZlt{}attributes\PYZgt{}}
          \PY{n+nt}{\PYZlt{}angle\PYZgt{}}0\PY{n+nt}{\PYZlt{}/angle\PYZgt{}}
          ...
        \PY{n+nt}{\PYZlt{}/attributes\PYZgt{}}
      \PY{n+nt}{\PYZlt{}/ViewSetup\PYZgt{}}
      ...
      \PY{n+nt}{\PYZlt{}Attributes} \PY{n+na}{name=}\PY{l+s}{\PYZdq{}angle\PYZdq{}}\PY{n+nt}{\PYZgt{}}
        \PY{n+nt}{\PYZlt{}Angle\PYZgt{}}
          \PY{n+nt}{\PYZlt{}id\PYZgt{}}0\PY{n+nt}{\PYZlt{}/id\PYZgt{}}
          \PY{n+nt}{\PYZlt{}name\PYZgt{}}45 degree\PY{n+nt}{\PYZlt{}/name\PYZgt{}}
        \PY{n+nt}{\PYZlt{}/Angle\PYZgt{}}
        ...
      \PY{n+nt}{\PYZlt{}/Attributes\PYZgt{}}
    \PY{n+nt}{\PYZlt{}/ViewSetups\PYZgt{}}
  \PY{n+nt}{\PYZlt{}/ViewRegistrations\PYZgt{}}
\PY{n+nt}{\PYZlt{}/SpimData\PYZgt{}}
\end{Verbatim}
\fvsettext
\noindent
An \Attributes element (lines 14--20) defines all attribute values for a given attribute |name|, in this case ``|angle|''.
Each of the values must at least have an \id, which is then used to associate an attribute value to a particular \ViewSetup.
This is exemplified in line 9, where the value with id |0| of the attribute named ``|angle|'' is referenced.

Adding a new type of attribute requires
\begin{enumerate}
  \item a Java class |A| that extends |Entity| (which means that it has an id) and represents an attribute value,
  \item a Java class that extends |XmlIoEntity<A>| and is annotated by a |@ViewSetupAttributeIo| annotation specifying the name of the attribute.
\end{enumerate}
For example, the above |angle| attribute is implemented by classes
\plset{language=java}
\fvsetlist
\begin{Verbatim}[commandchars=\\\{\}]
\PY{k+kd}{public} \PY{k+kd}{class} \PY{n+nc}{Angle} \PY{k+kd}{extends} \PY{n}{NamedEntity}
\PY{o}{\PYZob{}} \PY{o}{...}
\PY{o}{\PYZcb{}}
\end{Verbatim}
\noindent
and
\begin{Verbatim}[commandchars=\\\{\}]
\PY{n+nd}{@ViewSetupAttributeIo}\PY{o}{(}\PY{n}{name} \PY{o}{=} \PY{l+s}{\PYZdq{}angle\PYZdq{}}\PY{o}{,} \PY{n}{type} \PY{o}{=} \PY{n}{Angle}\PY{o}{.}\PY{n+na}{class}\PY{o}{)}
\PY{k+kd}{public} \PY{k+kd}{class} \PY{n+nc}{XmlIoAngle} \PY{k+kd}{extends} \PY{n}{XmlIoNamedEntity}\PY{o}{\PYZlt{}}\PY{n}{Angle}\PY{o}{\PYZgt{}}
\PY{o}{\PYZob{}} \PY{o}{...}
\PY{o}{\PYZcb{}}
\end{Verbatim}
\fvsettext
\noindent
The actual implementations are beyond the scope of this document.\footnote{We refer to
  \href{http://github.com/tpietzsch/spimdata}{\url{http://github.com/tpietzsch/spimdata}}
  for example attribute implementations, \eg, for angle.
}
The |@ViewSetupAttributeIo| annotation allows automatic discovery of the classes that handle particular attibutes.
Similar to Section~\ref{sec:ffcustomsections}, the XML reader/writer stores unhandled attributes as XML trees when reading files and puts them back into place when writing files.
This allows programmatic modification of datasets without understanding all attributes.
Again, we avoid losing information while also avoiding the need for every consumer to handle every custom attribute.

\chapter{User Guide}

\newcommand\button[1]{\textit{#1}}
\newcommand\key[1]{\texttt{#1}}
\newcommand\screenshotA[1]{\centerline{\includegraphics[width=.6\textwidth]{#1}}}
\newcommand\screenshotB[1]{\centerline{\includegraphics[width=.6\textwidth]{#1}}}
\newcommand\screenshotC[2]{\centerline{\includegraphics[width=.6\textwidth]{#1}}}

\section{Overview}

The \bdv is a re-slicing browser for terabyte-sized multi-view image sequences, for example multi-view light-sheet microscopy data.
Conceptually, the visualized data comprises multiple \emph{data sources} or \emph{setups}.
Each source provides one 3D image for each time-point (in the case of a time-lapse sequence).
For example, in a multi-angle SPIM sequence, each angle is a setup.
In a multi-angle, multi-channel SPIM sequence, each channel of each angle is a setup.

\Bdv comes with a custom data format that is is optimized for fast random access to very large data sets.
This permits browsing to any location within a multi-terabyte recording in a fraction of a second.
The file format is based on XML and HDF5~\cite{sup:hdf5} and is describe in the Supplementary Note~\supplFileFormatNumber.

This supplement is a slightly modified version of the \bdv User Guide on the Fiji wiki, \href{http://fiji.sc/BigDataViewer}{\url{http://fiji.sc/BigDataViewer}}.
In particular, we removed content which is redundant with Supplementary Note~\supplFileFormatNumber.
The User Guide describes the Fiji plugins that comprise \bdv, \ie, the viewer itself as well as plugins for importing and exporting data to/from our file format.

\section{Installation}
\Bdv is a Fiji plugin that is distributed via a Fiji update site.
You will need a recent version of Fiji which you can download from \href{http://fiji.sc}{\url{http://fiji.sc}}.
To install \bdv you need to enable its update site in the Fiji Updater.
Select \menu{Help > Update Fiji} from the Fiji menu to start the updater.
\\[2mm]
\screenshotB{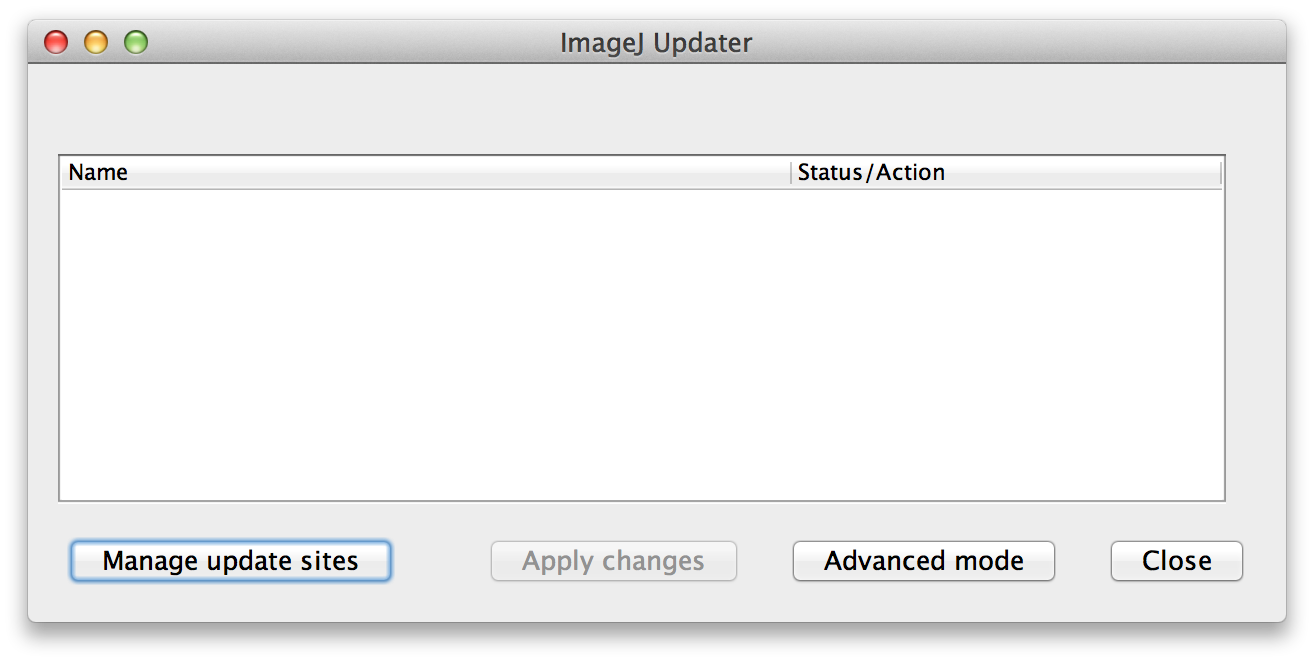}
\\
Click on \button{Manage update sites}. This brings up a dialog where you can activate additional update sites.
\\[2mm]
\screenshotB{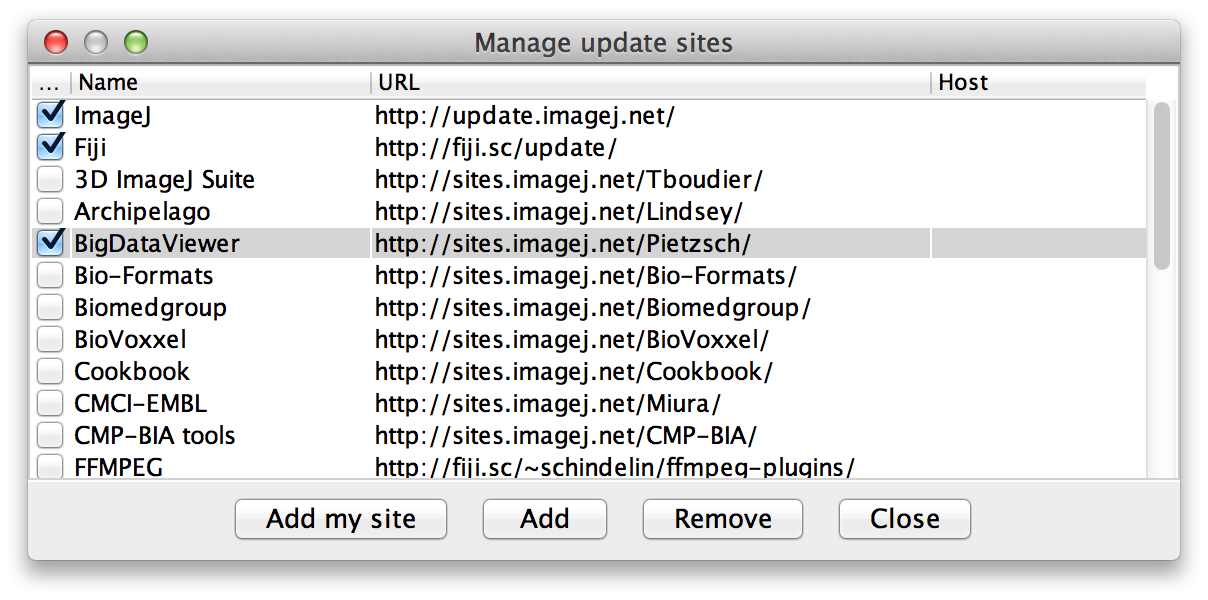}
\\
Activate the \bdv update site and \button{Close} the dialog.
Now you should see additional files appearing for download.
\\
\screenshotB{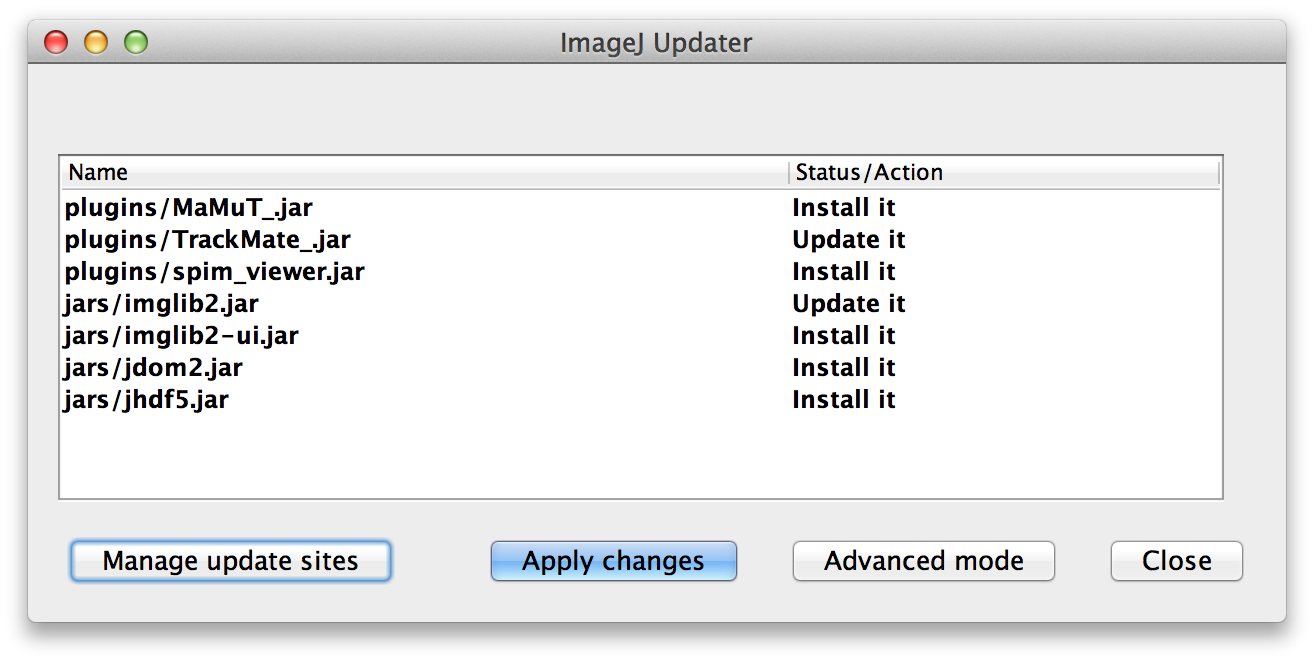}
\\
Click \button{Apply changes} and restart Fiji.

You should now have a sub-menu \menu{Plugins > BigDataViewer}.

\section{Usage}
To use the BigDataViewer we need some example dataset to browse.
You can download a small dataset from
\url{http://fly.mpi-cbg.de/~pietzsch/bdv-example/}, comprising two views and three time-points.
This is an excerpt of a 6 angle 715 time-point sequence of \emph{drosophila melanogaster} embryonal development, imaged with a Zeiss Lightsheet Z.1.
Download both the XML and the HDF5 file and place them somewhere next to each other.

Alternatively, you can create a dataset by exporting your own data as described below.

\subsection{Opening Dataset}
To start \bdv, select \menu{Plugins > BigDataViewer > Open XML/HDF5} from the Fiji menu.
This brings up a file open dialog. Open the XML file of your test dataset.

\subsection{Basic Navigation}
You should see something like this:
\\[2mm]
\screenshotA{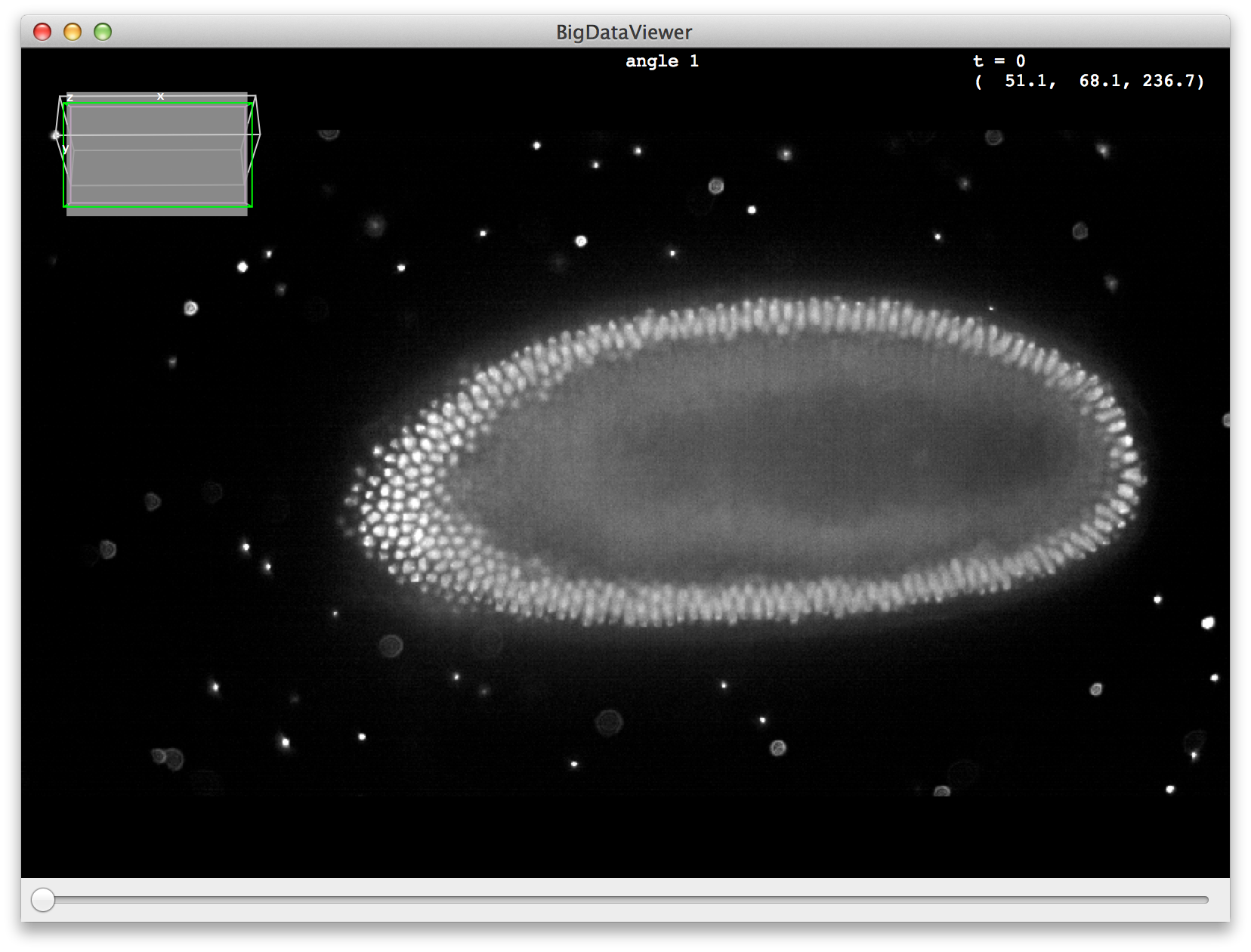}
\\
On startup, the middle slice of the first source (angle) is shown.
You can browse the stack using the keyboard or the mouse.
To get started, try the following:
\begin{itemize}
  \item Use the mouse-wheel or \keys{<} and \keys{>} keys to scroll through z slices.
  \item \keys{right-click + drag} anywhere on the canvas to translate the image.
  \item Use \keys{\ctrl + \shift + mouse-wheel}, or \keys{\arrowkeyup} and \keys{\arrowkeydown} keys to zoom in and out.
  \item \keys{left-click + drag} anywhere on the canvas to rotate (reslice) the image.
\end{itemize}

The following table shows the available navigation commands using the mouse:
\vspace{-2mm}
\newcommand{\specialcell}[2][c]{%
  \begin{tabular}[#1]{@{}l@{}}#2\end{tabular}}
\newcommand\tablecell[2][&]{#2 #1}

{
\renewcommand{\arraystretch}{1.5}
\begin{center}
\begin{tabular}{ | p{.3\textwidth - 2\tabcolsep} | p{.7\textwidth - 2\tabcolsep} |}
    \hline
  \tablecell{\keys{left-click + drag}}
  \tablecell[\\]{Rotate (pan and tilt) around the point where the mouse was clicked.}
  \hline
  \tablecell{\specialcell{\keys{right-click + drag} or\\ \keys{middle-click + drag}}}
  \tablecell[\\]{Translate in the XY-plane.}
  \hline
  \tablecell{\keys{mouse-wheel}}
  \tablecell[\\]{Move along the z-axis.}
  \hline
  \tablecell{\specialcell{\keys{\cmd + mouse-wheel} or\\ \keys{\shift + \ctrl + mouse-wheel}}}
  \tablecell[\\]{Zoom in and out.}
  \hline
\end{tabular}
\end{center}
}

The following table shows the available navigation commands using keyboard shortcuts:
{
\renewcommand{\arraystretch}{1.5}
\begin{center}
\begin{tabular}{ | p{.3\textwidth - 2\tabcolsep} | p{.7\textwidth - 2\tabcolsep} |}
    \hline
  \tablecell{\keys{X}, \keys{Y}, \keys{Z}}
  \tablecell[\\]{Select keyboard rotation axis.}
  \hline
  \tablecell{\keys{\arrowkeyleft}, \keys{\arrowkeyright}}
  \tablecell[\\]{Rotate clockwise or counter-clockwise around the choosen rotation axis.}
  \hline
  \tablecell{\keys{\arrowkeyup}, \keys{\arrowkeydown}}
  \tablecell[\\]{Zoom in or out.}
  \hline
  \tablecell{\keys{,}, \keys{.}}
  \tablecell[\\]{Move forward or backward along the Z-axis.}
  \hline
  \tablecell{\keys{\shift + X}}
  \tablecell[\\]{Rotate to the ZY-plane of the current source. (Look along the X-axis of the current source.)}
  \hline
  \tablecell{\keys{\shift + Y} or \keys{\shift + A}}
  \tablecell[\\]{Rotate to the XZ-plane of the current source. (Look along the Y-axis of the current source.)}
  \hline
  \tablecell{\keys{\shift + Z}}
  \tablecell[\\]{Rotate to the XY-plane of the current source. (Look along the Z-axis of the current source.)}
  \hline
  \tablecell{\keys{[} or \keys{N}}
  \tablecell[\\]{Move to previous timepoint.}
  \hline
  \tablecell{\keys{]} or \keys{M}}
  \tablecell[\\]{Move to next timepoint.}
  \hline
\end{tabular}
\end{center}
}

\noindent
For all navigation commands you can hold \keys{\shift} to rotate and browse
$10\times$ faster, or hold \keys{\ctrl} to rotate and browse
$10\times$ slower.
For example, \keys{\arrowkeyleft} rotates by
$1^\circ$ clockwise, while \keys{\shift + \arrowkeyleft} rotates by
$10^\circ$, and \keys{\ctrl + \arrowkeyleft} rotates by
$0.1^\circ$.

The axis-rotation commands (\eg, \keys{\shift + X}) rotate around the current mouse location.
That is, if you press \keys{\shift + X}, the view will pivot such that you see a ZY-slice through the dataset (you look along the X-axis).
The point under the mouse will stay fixed, \ie, the view will be a ZY-slice through that point.

\subsection{Interpolation Mode}
Using \keys{I} you can switch between nearest-neighbor and trilinear interpolation schemes.
The difference is clearly visible when you zoom in such that individual source pixels are visible.
\\
\screenshotA{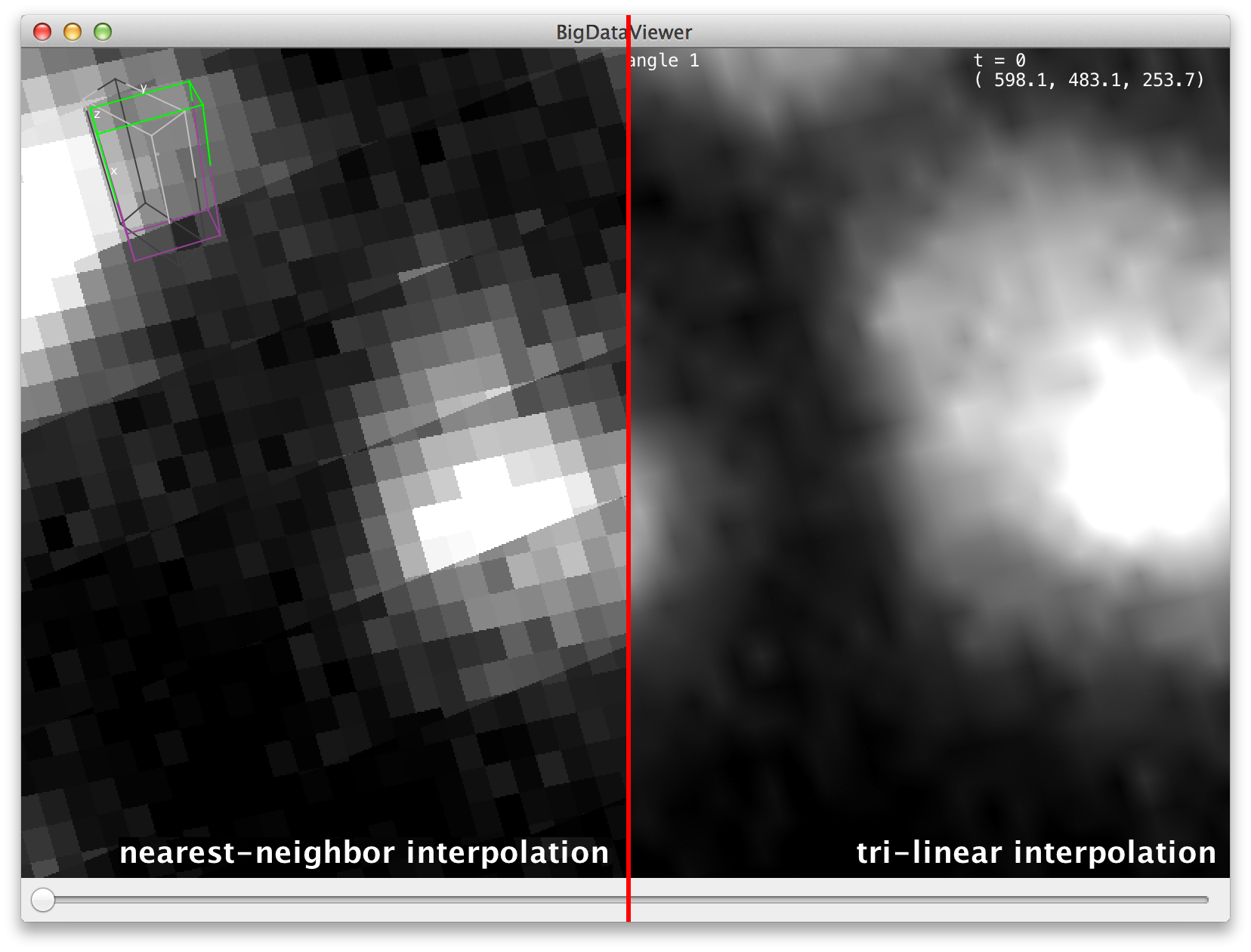}
\\
Trilinear interpolation results in smoother images but is a bit more expensive computationally.
Nearest-neighbor is faster but looks more pixelated.

\subsection{Displaying Multiple Sources}
\Bdv datasets typically contain more than one source.
For a SPIM sequence one usually has multiple angles and possibly fused and deconvoled data on top.

Select \menu{Settings > Visibility \& Grouping} from the \bdv menu to bring up a dialog to control source visibility.
You can also bring up this dialog by the shortcut \keys{F6}.
\\[2mm]
\screenshotB{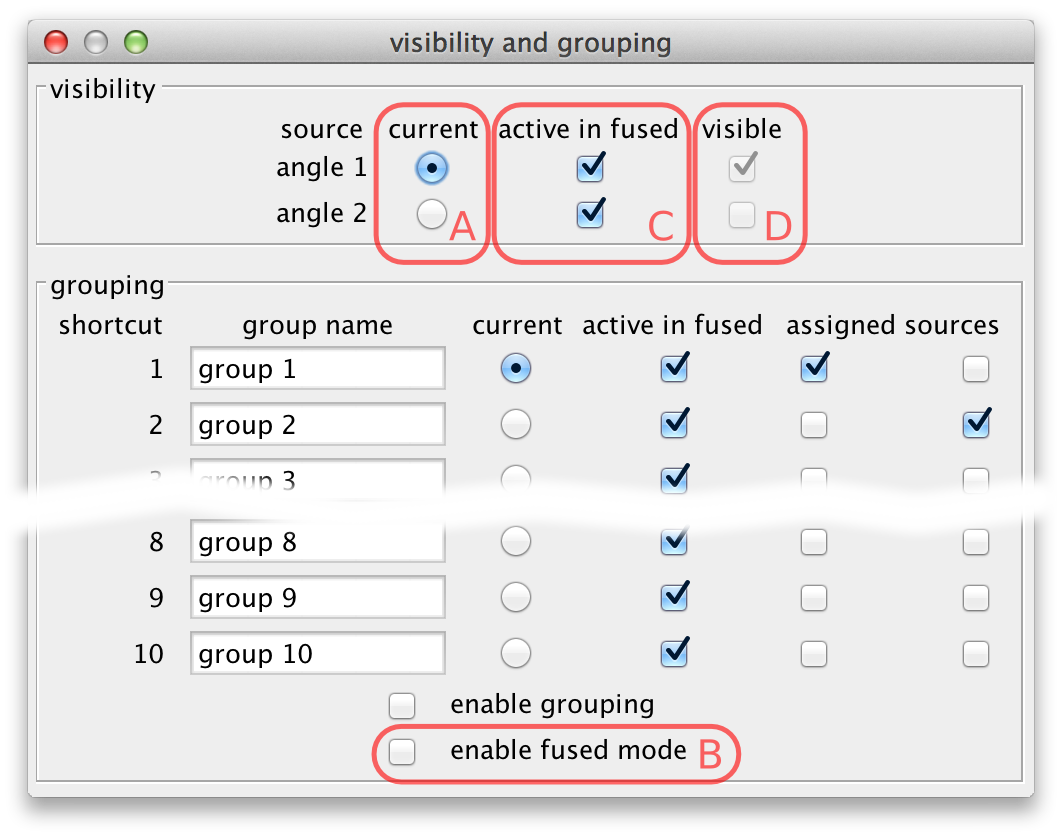}
\\
Using the current source checkboxes (A in the figure above), you can switch between available sources.
The first ten sources can also be made current by the number keys \keys{1} through \keys{0} in the main \bdv window.

To view multiple sources overlaid at the same time, switch to \emph{fused mode} using the checkbox (B).
You can also switch between normal and fused mode using the shortcut \keys{F} in the main window.
In fused mode individual sources can be turned on and off using the checkboxes (C) or shortcuts \keys{\shift + 1} through \keys{\shift + 0} in the main window.

Whether in normal or fused mode, the (unselectable) boxes (D) provide feedback on which sources are actually currently displayed.
Also the main window provides feedback:
\\[2mm]
\screenshotB{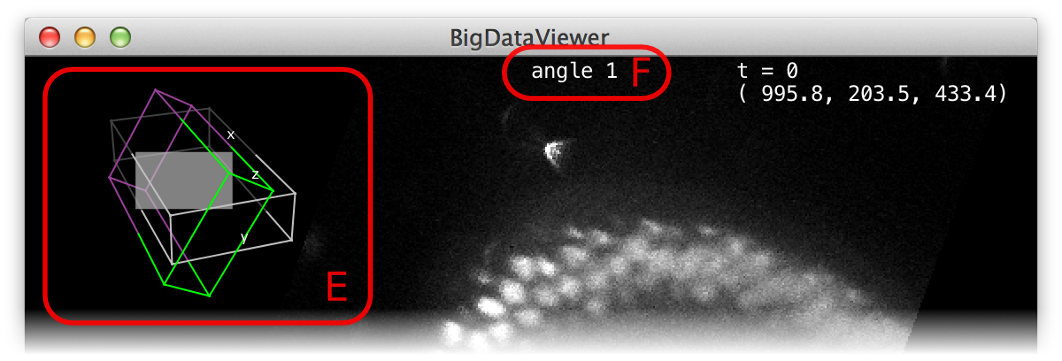}
\\
In the top-left corner an overview of the dataset is displayed (E).
Visible sources are displayed as green/magenta wireframe boxes, invisible sources are displayed as grey wireframe boxes.
The dimensions of the boxes illustrate the size of the source images. The filled grey rectangle illustrates the screen area, \ie, the portion of the currently displayed slice.
For the visible sources, the part that is in front of the screen is green, the part that is behind the screen is magenta.

At the top of the window, the name of the current source is shown (F).

Note, that also in fused mode there is always a \emph{current source}, although this source may not even be visible.
Commands such as \keys{\shift + X} (rotate to ZY-plane) refer to the local coordinate system of the current source.

\subsection{Grouping Sources}
Often there are sets of sources for which visibility is logically related.
For example, in a multi-angle, multi-channel SPIM sequence, you will frequently want to see all channels of a given angle, or all angles of a given channel.
If your dataset contains deconvolved data, you may want to see either all raw angles overlaid, or the deconvolved view, respectively.
You want to be able to quickly switch between those two views.
Turning individual sources on and off becomes tedious in these situations.
Therefore, sources can be organized into \emph{groups}.
All sources of a \emph{group} can be activated or deactivated at once.

Source grouping is handled in the visibility and grouping dialog, too
(menu \menu{Settings > Visibility \& Grouping} or shortcut \keys{F6}).
\\
\screenshotB{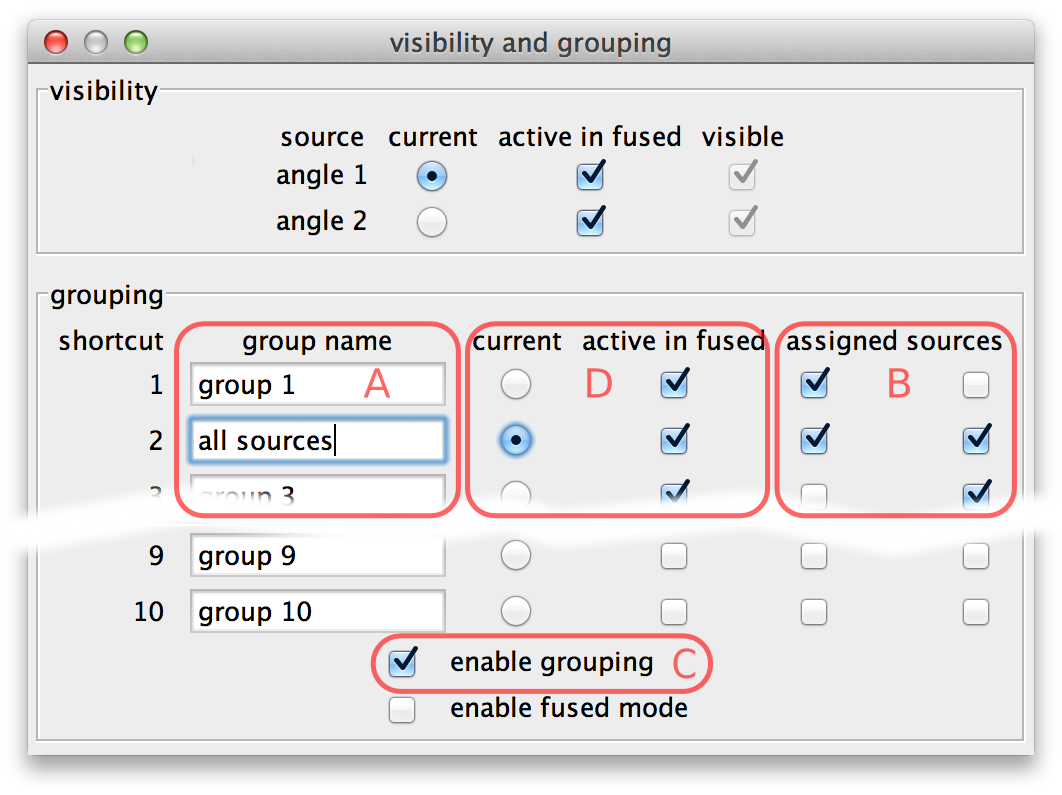}
\\
The lower half of the dialog is dedicated to grouping.
There are 10 groups available.
They are named ``group 1'' through ``group 10'' initially, but the names can be edited (A).

Sources can be assigned to groups using the checkboxes (B).
In every line, there are as many checkboxes as there are sources.
Sources corresponding to active checkboxes are assigned to the respective group.
For example, in the above screenshot there are two sources and therefore two ``assigned sources'' checkboxes per line
The first source is assigned to groups 1 and 2, the second source is assigned to groups 2 and 3.
Group 2 has been renamed to ``all sources''.

\emph{Grouping} can be turned on and off by the checkbox (C) or by using the shortcut \keys{G} in the main window.
If grouping is enabled, groups take the role of individual sources:
There is one \emph{current group} which is visible in normal mode (all individual sources that are part of this group are overlaid).
Groups can be activated or deactivated to determine visibility in fused mode (all individual sources that are part of at least one active group are overlaid).

Groups can be made current and made active or inactive using the checkboxes (D).
Also, if grouping is enabled the number key shortcuts in the main \bdv window act on groups instead of individual sources.
That is, groups 1 through 10 can be made current by keys \keys{1} through \keys{0}.
Similarly, shortcuts \keys{\shift + 1} through \keys{\shift + 0} in the main window activate or deactivate groups 1 through 10 for visibility in fused mode.

If grouping is enabled, the name of the current group is shown at the top of the main window.
\\[2mm]
\screenshotB{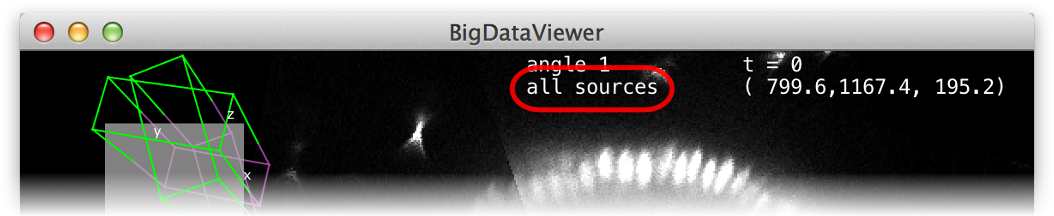}

\subsection{Adjusting Brightness and Color}
To change the brightness, contrast, or color of particular sources select \menu{Setting > Brightness \& Color} or press the shortcut \keys{S}.
This brings up the brightness and color settings dialog.
\\[2mm]
\screenshotB{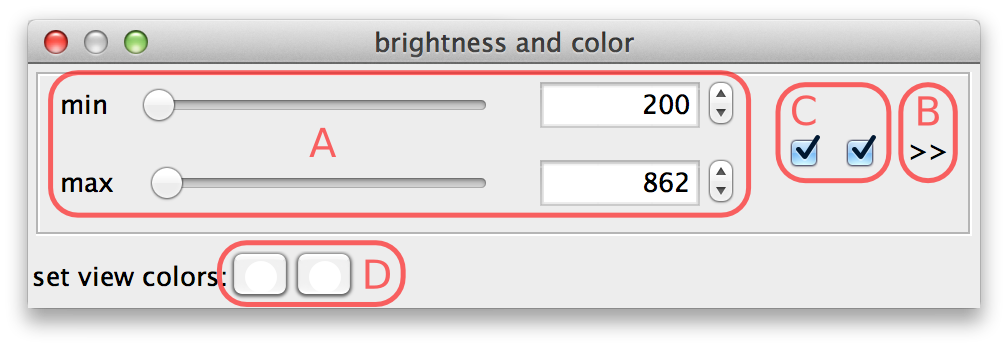}
\\
The \emph{min} and \emph{max} sliders (A) can be used to adjust the brightness and contrast.
They represent minimum and maximum source values that are mapped to the display range.
For the screenshot above, this means that source intensity 200 (and everything below) is mapped to black. Source intensity 862 (and everything above) is mapped to white.

When a new dataset is opened, \Bdv tries to estimate good initial \emph{min} and \emph{max} settings by looking at the first image of the dataset.

\Bdv datasets are currently always stored with 16 bits per pixel, however the data does not always exploit the full value range 0 \dots 65535.
The example drosophila dataset uses values in the range of perhaps 0 \dots 1000, except for the much brighter fiducial beads around the specimen.
The \emph{min} and \emph{max} sliders in this case are a bit fiddly to use, because they span the full 16 bit range with the interesting region squeezed into the first few pixels.
This can be remedied by adjusting the range of the sliders.
For this, click on the $>>$ dialog button (B).
This shows two additional input fields, where the range of the sliders can be adjusted.
In the following screenshot, the leftmost value of the slider range has been set to 0 and the rightmost value to 2000, making the sliders much more useful.
\\[2mm]
\screenshotB{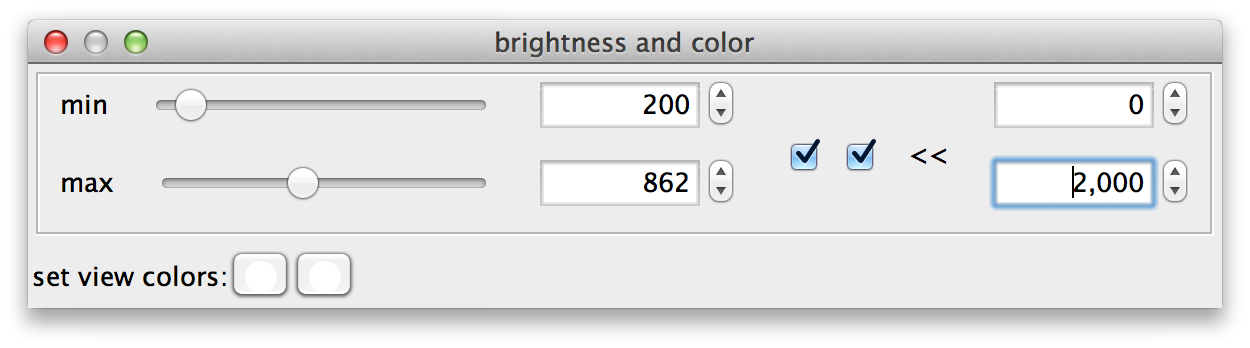}
\\
So far, all sources share the same \emph{min} and \emph{max} settings.
However, these can also be adjusted for each individual source or for groups of sources.
The checkboxes (C) assign sources to \emph{min-max-groups}.
There is one checkbox per source.
In the example drosophila dataset there are two sources, therefore there are two checkboxes.
The active checkboxes indicate for which sources the \emph{min} and \emph{max} values apply.

If you uncheck one of the sources, it will move to its own new \emph{min-max-group}.
Now you can adjust the values for each source individually.
The sliders of new group are initialized as a copy of the old group.
\\[2mm]
\screenshotB{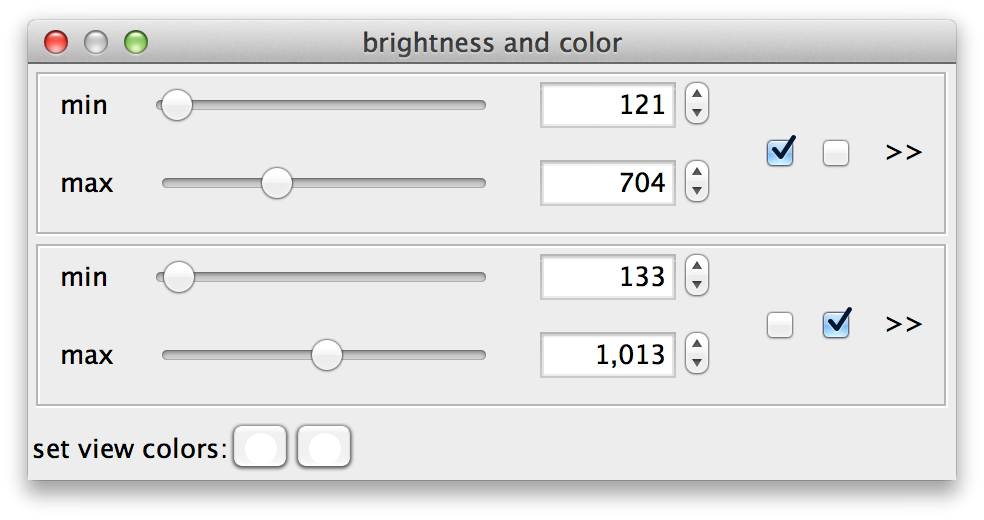}
\\
Sources can be assigned to \emph{min-max-groups} by checking/unchecking the checkboxes.
The rule is that every source is always assigned to exactly one min-max-group.
Thus, if you activate an unchecked source in a min-max-group, this will remove the source from its previous min-max-group and add it to the new one.
Unchecking a source will remove it from its min-max-group and move it to a new one.
Min-max-groups that become empty are removed.
To go back to a single min-max-group in the example, simply move all sources to the same group.

Finally, at the bottom of the dialog (D) colors can be assigned to sources.
There is one color button per source (two in the example).
Clicking a button brings up a color dialog, where you can choose a color for that particular source.
In the following screenshot, the sources have been colored magenta and green.
\\
\centerline{\includegraphics[width=.55\textwidth]{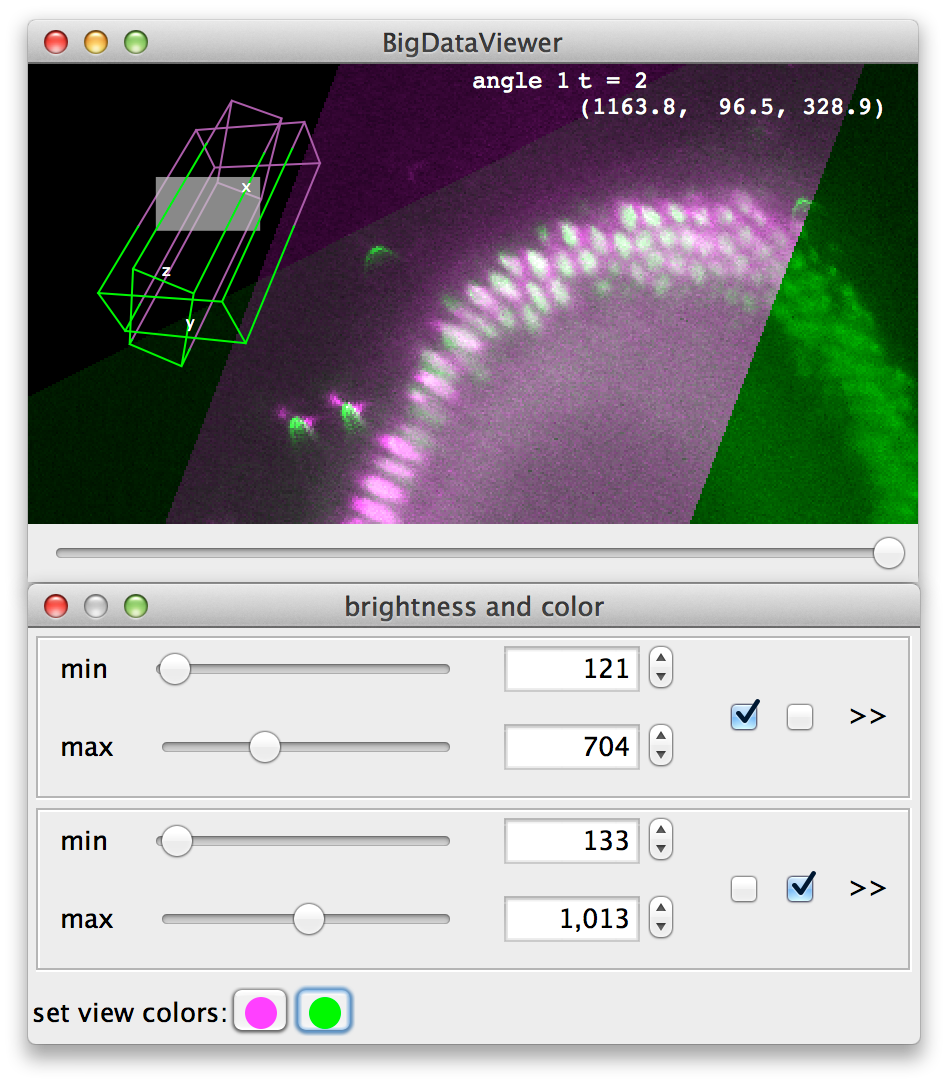}}

\subsection{Bookmarking Locations and Orientations}
\Bdv allows to bookmark the current view.
You can set bookmarks for interesting views or particular details of your dataset to easily navigate back to those views later.

Each bookmark has an assigned shortcut key, \ie, you can have bookmarks ``a'', ''A'', ''b'', \dots, ``1'', ``2'', \etc.
To set a bookmark for the current view, press \keys{\shift + B} and then the shortcut you want to use for the bookmark.
To recall bookmark, press \keys{B} and then the shortcut of the bookmark.

\Bdv provides visual feedback for setting and recalling bookmarks.
When you press \keys{\shift + B}, the message ``\textbf{set bookmark:}'' appears in the lower right corner of the main window, prompting to press the bookmark shortcut next.
\\[2mm]
\screenshotB{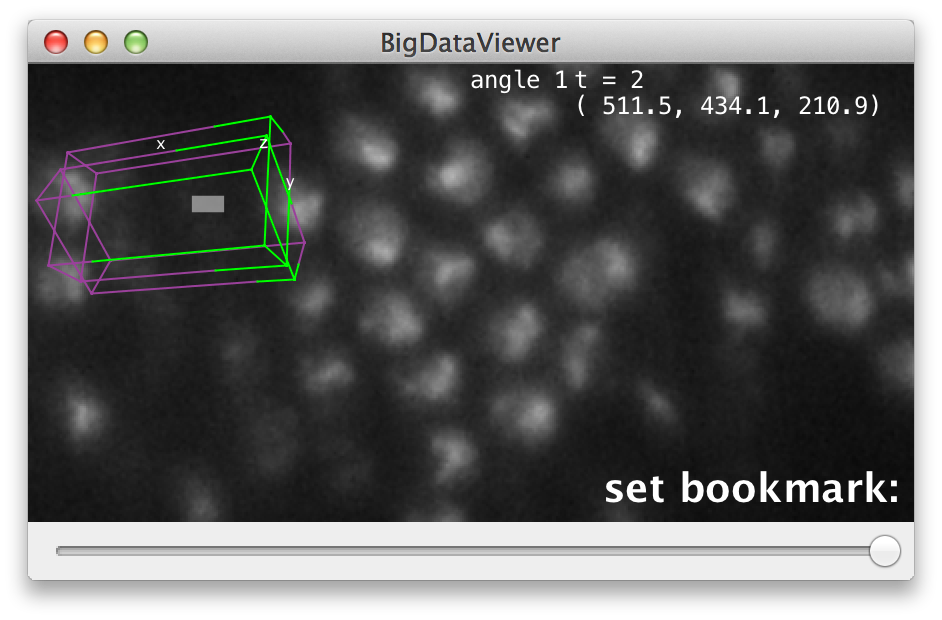}
\\
Now press the key you want to use as a shortcut, for example \keys{A}.
The promt message will change to ``\textbf{set bookmark: a}'' indicating that you have set a bookmark with shortcut \keys{A}.
Instead of pressing a shortcut key you can abort using \keys{esc}.

Similarly, when you press \keys{B} to recall a bookmark, the prompt message ``\textbf{go to bookmark:}'' appears.
Now press the shortcut of the bookmark you want to recall, for example \keys{A}.
The promt message will change to ``\textbf{go to bookmark: a}'' and the view will move to the bookmarked location.
Instead of pressing a shortcut key you can abort using \keys{esc}.

Note, that bookmark shortcuts are case-sensitive, \ie, \keys{A} and \keys{\shift + A} refer to distinct bookmarks ``a'' and ``A'' respectively.

The bookmarking mechanism can also be used to bookmark and recall orientations.
Press \keys{O} and then a bookmark shortcut to recall only the orientation of that bookmark.
This rotates the view into the rotation of the bookmarked view (but does not zoom or translate to the bookmarked location).
The rotation is around the current mouse location (\ie, the point under the mouse stays fixed).

\subsection{Loading and Saving Settings}
Organizing sources into groups, assigning appropriate colors, adjusting brightness correctly, and bookmarking interesting locations is work that you do not want to repeat over and over every time you re-open a dataset.
Therefore, \Bdv allows to save and load these settings.

Select \menu{File > Save settings} from the menu to store settings to an XML file, and \menu{File > Load settings} to load them from an XML file.

When a dataset is opened, \bdv automatically loads an appropriately named settings file if it is present.
This settings file must be in the same directory as the dataset's XML file, and have the same filename with \emph{.settings} appended.
For example, if the dataset's XML file is named \emph{drosophila.xml}, the settings file must be named \emph{drosophila.settings.xml}.
(If you select \menu{File > Save settings}, this filename is already suggested in the Save File dialog.)

Settings files assume that a specific number of sources are present, therefore settings are usually not compatible across different datasets.

\section{Opening \Bdv Datasets as ImageJ Stacks}
\Bdv may be great for looking at your data, but what if you want to apply other ImageJ algorithms or plugins to the images?
You can open individual images from a dataset as ImageJ stacks using \menu{File > Import > BigDataViewer...} from the Fiji menu.
\\[2mm]
\screenshotB{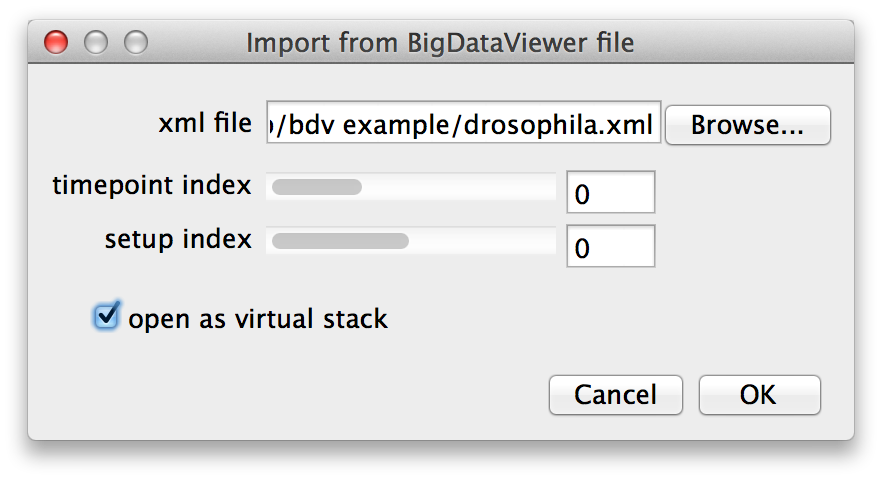}
\\
Select the XML file of a dataset, then choose the time-point and source (setup) index of the image you want to open.
If you enable the \emph{``open as virtual stack''} checkbox the image will open as an ImageJ \emph{virtual stack}.
This means that the opened image is backed by \bdv's cache and slices are loaded on demand.
Without \emph{``open as virtual stack''}, the full image will be loaded into memory.
Virtual stacks will open a bit faster but switching between slices may be less instantaneous.

Note that the import function is macro-recordable.
Thus, you can make use of it to batch-process images from \bdv datasets.


%
%
\section{Exporting Datasets for the \Bdv}
BigDataViewer uses a custom file-format that is optimized for fast arbitrary re-slicing at various scales.
This file format is build on open standards XML\cite{sup:xml} and HDF5\cite{sup:hdf5}, where HDF5 is used to store image volumes\footnote{
    Actually, we support several ways to store the image volumes besides HDF5.
    For example, the volume data can be provided by a web service for remote access.
    However, the Fiji plugins always export to HDF5.}
  and XML is used to store meta-data.
The format is explained in detail in Supplementary Note \supplFileFormatNumber --
  we recommend to read at least the overview in Section~2 of that Note for some background, rationale, and terminology that will be helpful in the following.

\subsection{Exporting from ImageJ Stacks}
You can export any dataset to \bdv format by opening it as a stack in Fiji and then selecting \menu{Plugins > BigDataViewer > Export Current Image as XML/HDF5} from the Fiji menu.
If the image has multiple channels, each channel will become one \emph{setup} in the exported dataset.
If the image has multiple frames, each frame will become on \emph{timepoint} in the exported dataset.
Of course, you may export from virtual stacks if your data is too big to fit into memory.

To get started, let's open one of the ImageJ sample images by \menu{File > Open Samples > T1 Head (2.4M, 16-bits)}.
Selecting \menu{Plugins > BigDataViewer > Export Current Image as XML/HDF5} brings up the following dialog.
\\[2mm]
\screenshotB{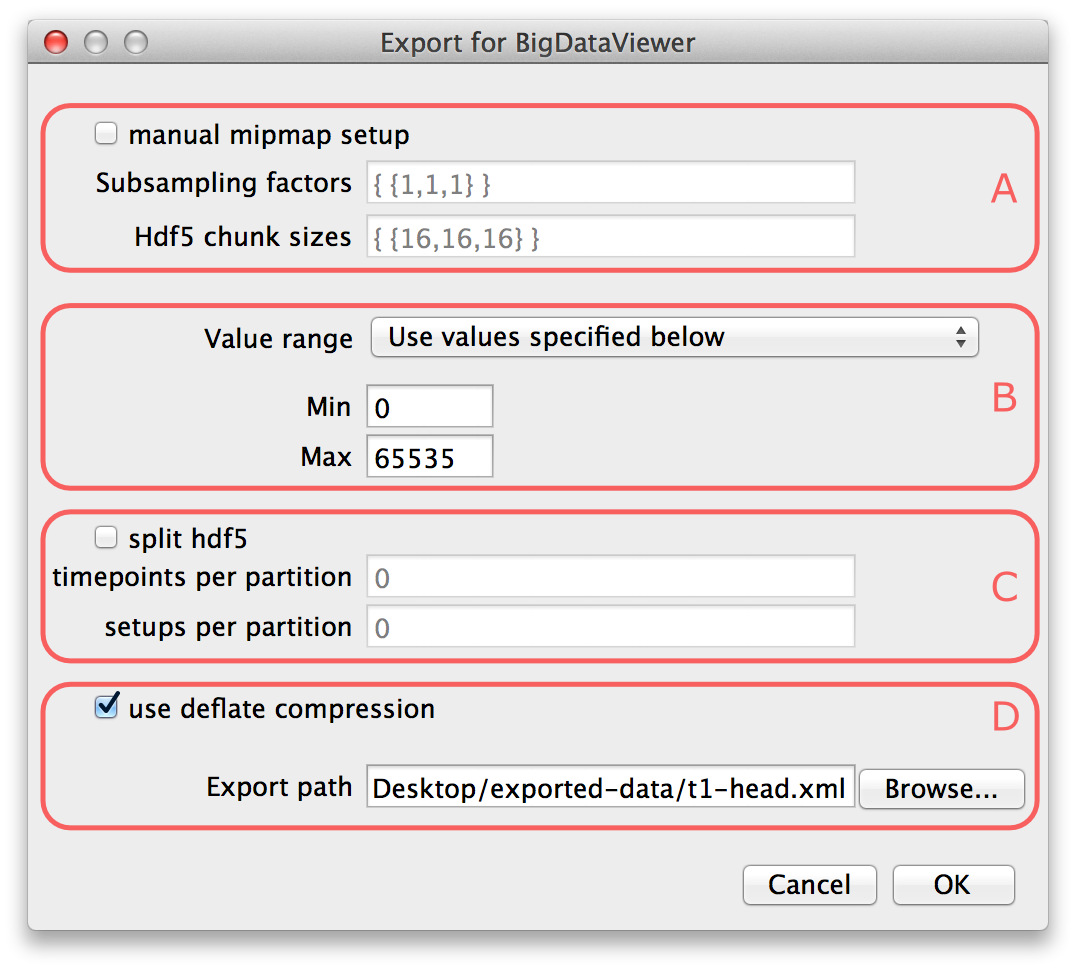}
\\
Parts (A) and (C) of the dialog are optional, so we will explain (B) and (D) first.

At the bottom of the dialog (D), the export path is defined.
Specify the path of the XML file to which you want to export the dataset.
The HDF5 file for the dataset will be placed into the same directory under the same name with extension ``.h5''.
If the \emph{``use deflate compression''} checkbox is enabled, the image data will be compressed using HDF5 built-in DEFLATE compression.
We recommend to use this option.
It will usually reduce the file size to about 50\% with respect to uncompressed image size.
The performance impact of decompression when browsing the dataset is negligible.

In part (B) of the dialog the \emph{value range} of the image must be specified.
\Bdv always stores images with 16-bit precision currently, while the image you want to export is not necessarily 16-bit.
The value range defines the minimum and maximum of the image you want to export.
This is mapped to the 16-bit range for export.
\Ie, the minimum of the value range will be mapped to the minimum of the unsigned 16-bit range (0).
The maximum of the value range will be mapped to the maximum of the unsigned 16-bit range (65535).
In the drop-down menu you can select one the following options to specify how the value range should be determined:
\begin{itemize}
  \item \emph{``Use ImageJ's current min/max setting''}
    The minimum and maximum set in ImageJ's Brightness\&Contrast are used.
    Note, that image intensities outside that range will be clipped to the minimum or maximum, respectively.
  \item \emph{``Compute min/max of the (hyper-)stack''}
    Compute the minimum and maximum of the stack and use these.
    Note, that this may take some time to compute because it requires to look at all pixels of the stack you want to export.
  \item \emph{``Use values specified below''}
    Use the values specified in the \emph{Min} and \emph{Max} fields (B) of the export dialog.
    Note, that image intensities outside that range will be clipped to the minimum or maximum, respectively.
\end{itemize}

After you have specified the value range and selected and export path, press \emph{OK} to export the dataset.
Messages about the progress of the operation are displayed in the ImageJ Log window.
\\[2mm]
\screenshotB{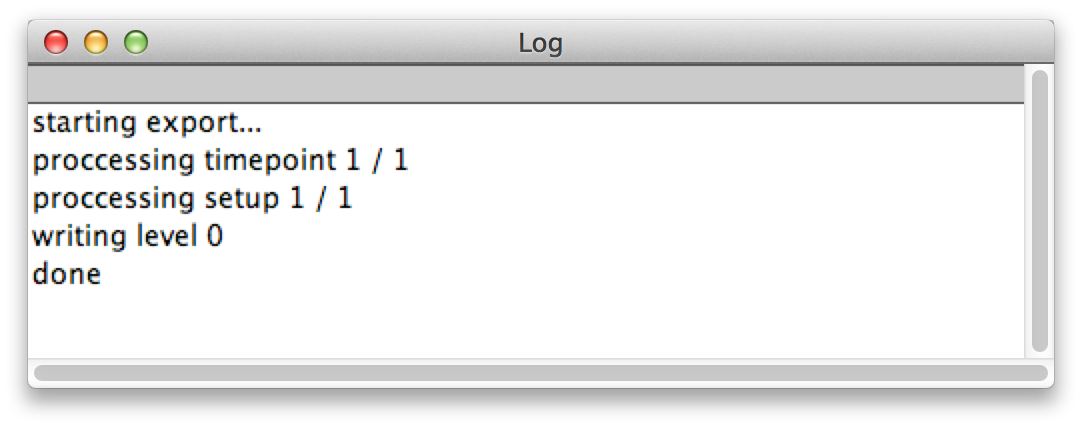}
\\
When the export is done you can browse the dataset in the \bdv by opening the exported XML file.

The optional parts (A) and (C) of the export dialog provide further options to customize the export.
If the checkbox \emph{``manual mipmap setup''} is enabled, you can customize the multi-resolution mipmap pyramid which stores your image stacks.
You can specify the number of resolution levels used, and their respective down-scaling factors, as well as the chunk sizes into which each resolution level is subdivided.

The \emph{``Subsampling factors''} field specifies a comma-separated list of resolution levels, formatted as \texttt{\{level, \dots, level\}}.
Each \texttt{level} is a list of subsampling factors in $X, Y, Z$, formatted as \texttt{\{x-scale, y-scale, z-scale\}}.
For example consider the specification \texttt{\{\{1,1,1\}, \{2,2,1\}, \{4,4,2\}\}}.
This will create a resolution pyramid with three levels.
The first level is the full resolution -- it is scaled by factors $1$ in all axes.
The second level is down-scaled by factors $2, 2, 1$ in $X, Y, Z$ respectively.
So it has half the resolution in $X$ and $Y$, but full resolution in $Z$.
The third level has half the resolution of the second in all axes, \ie, it is down-scaled by factors $4, 4, 2$ in $X, Y, Z$ respectively.
Note, that you should always order levels by decreasing resolution like this.
Also note, that in the above example we down-scale by different factors in different axes.
One may want to do this if the resolution of the dataset is anisotropic.
Then it is advisable to first downscale only in the higher-resolved axes until approximately isotropic resolution is reached.

The \emph{``Hdf5 chunk sizes''} specifies the chunk sizes into which data on each resolution level is sub-divided.
This is again formatted as \texttt{\{level, \dots, level\}}, with the same number of levels as supplied for the \emph{Subsampling factors}.
Each \texttt{level} is a list of sizes in $X, Y, Z$, formatted as \texttt{\{x-size, y-size, z-size\}}.
For example consider the specification \texttt{\{\{16,16,16\}, \{16,16,16\}, \{16,16,16\}\}}.
This will sub-divide each resolution level into chunks of $16\times 16\ times 16$ pixels.

It is usually not recommended to specify subsampling factors and chunk sizes manually.
When browsing a dataset, the mipmap setup determines the loading speed and therefore the perceived speed of browsing to data that is not cached.
With \emph{``manual mipmap setup''} turned off, reasonable values will be determined automatically depending on the resolution and anisotropy of your dataset.

Finally, in part (C) of the export dialog, you may choose to split your dataset into multiple HDF5 files.
This is useful in particular for very large datasets.
For example when moving the data to a different computer, it may be cumbersome to have it sitting in a single 10TB file.
If the checkbox \emph{``split hdf5''} is enabled the dataset will be split into multiple HDF5 \emph{partition} files.
The dataset can be split along the \emph{timepoint} and \emph{setup} dimensions.
Specify the number of \emph{timepoints per partition} and \emph{setups per partition} in the respective input fields.

For example, assume your dataset has 4 setups and 10 timepoints.
Setting \emph{timepoints per partition = 4} and \emph{setups per partition = 2} will result in 4 HDF5 partitions:
\begin{itemize}
  \item setups 1 and 2 of timepoints 1 through 5,
  \item setups 3 and 4 of timepoints 1 through 5,
  \item setups 1 and 2 of timepoints 6 through 10, and
  \item setups 3 and 4 of timepoints 6 through 10.
\end{itemize}
Setting \emph{timepoints per partition = 0} or \emph{setups per partition = 0} means that the dataset is not split in the respective dimension.

Note, that splitting into multiple HDF5 files is transparent from the viewer side.
There is still only one XML file that gathers all the partitions files into one dataset.

\subsection{Integration with Fiji's SPIMage Processing Tools}

\Bdv seamlessly integrates with the ``Multiview Reconstruction'' plugins that Fiji offers for registration and reconstruction of lightsheet microscopy data.
Recent versions of these tools build on the same XML format as \bdv itself.
In addition to HDF5, ``Multiview reconstruction'' supports a backend for datasets that store individual views as TIFF files,
  because unprocessed data from lightsheet microscopes is often available in this format.

In principle, \bdv is able to display a TIFF dataset as is.
However, for quick navigation this is not the ideal format:
When navigating to a new timepoint, \bdv needs to load all TIFF files of that timepoint into memory, suffering a delay of tens of seconcds.
Therefore, it is beneficial to convert the TIFF dataset to HDF5.
Note, that this can be one at any point of the processing pipeline (\ie, before registration, after registration, after multiview fusion or deconvolution, \etc)
  because the ``Multiview reconstruction'' plugins can operate on HDF5 datasets as well.

A discussion of the ``Multiview reconstruction'' plugins is beyond the scope of this document.
We assume that the used has already created an XML/TIFF dataset, and refer to the description on the Fiji wiki,
  \href{http://fiji.sc/Multiview-Reconstruction}{\url{http://fiji.sc/Multiview-Reconstruction}}, for details.

To convert the dataset to HDF5, select \menu{Plugins > Multiview Reconstruction > Resave > As HDF5} form the Fiji menu.
This brings up the following dialog.
\\[2mm]
\screenshotB{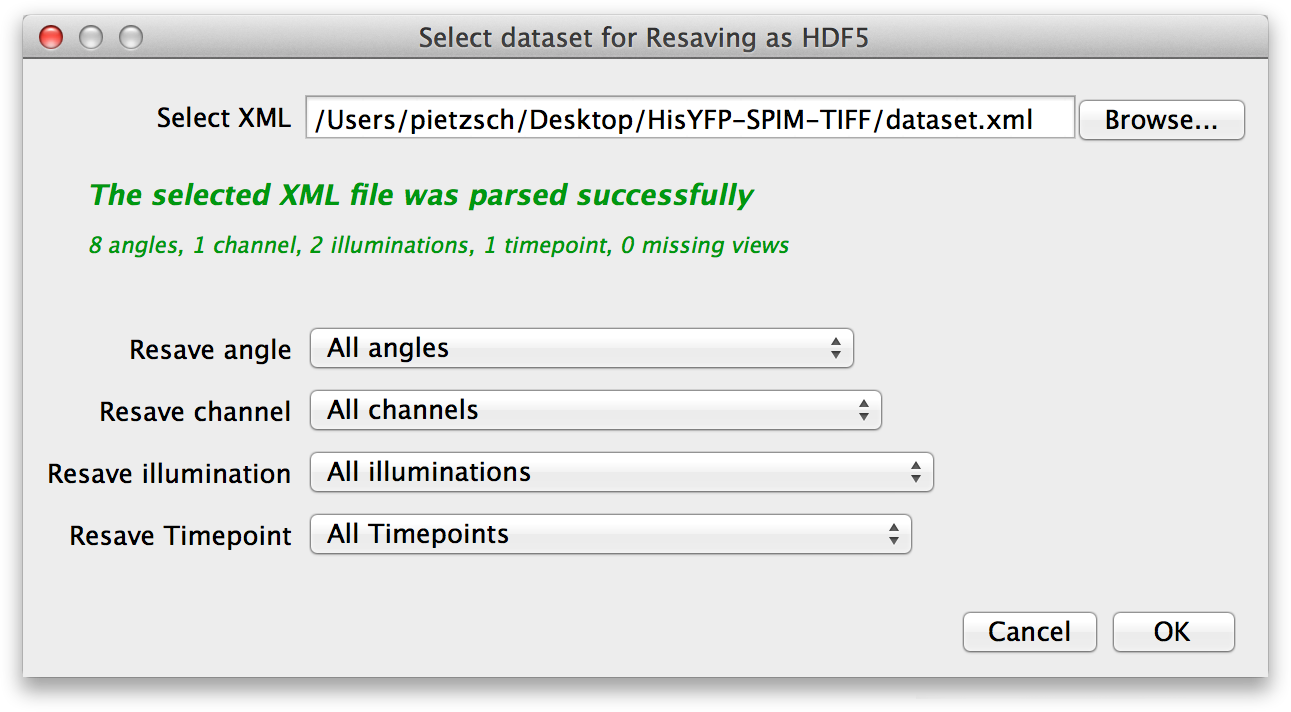}
\\
At the top of the dialog, select the XML file of the dataset you want to convert to HDF5.
In the lower part of the dialog, you can select which parts of the dataset you want to resave.
For example, assume that the source dataset contains the raw image stacks from the microscope, as well as deconvolved versions.
You might decide that you do not need the raw data as HDF5, so you can select only the deconvolved channels.
Once you have determined what you want to convert press \emph{OK}.

This brings up the next dialog, in which you need to specify the export path and options.
\\[2mm]
\screenshotB{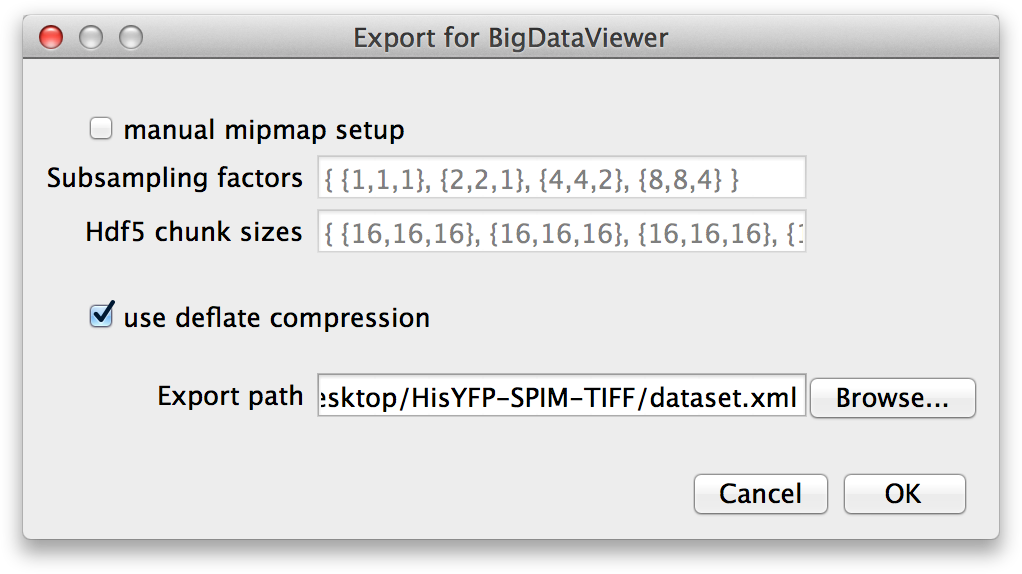}
\\
These parameters are the same as discussed in the previous section:
If you want to specify custom mipmap settings, you can do so in the top part of the dialog.
Below that, choose whether you want to compress the image data.
For the export path, specify the XML file to which you want to export the dataset.
Press \emph{OK} to start the export.

\chapter{Software Architecture}

\noindent
The architecture of \bdv separates data access, caching, and visualization into cleanly delimited, well-defined components.
To achieve this we build on ImgLib2~\cite{sup:imglib2}, a generic Java image processing library that serves three important purposes.

First, our cache-backed image data structures are built on top of ImgLib2's |CellImg|.
This is an image container that represents multi-dimensional images by splitting them into smaller chunks (|Cell|s).
We augment this functionality by loading and caching of chunks on demand.
With this, the image data is not required to be completely held in memory.
Second, our rendering algorithm relies on the facilities that ImgLib2 provides for virtualized pixel access and transformations,
  as explained in Supplementary Note~\supplRenderingNumber.
Finally, component boundaries and data exchange between components is defined in terms of ImgLib2 interfaces.

\newcommand\Component[1]{\textbf{#1}}
\newcommand\MetaData{\Component{MetaData}\xspace}
\newcommand\Source{\Component{Source}\xspace}
\newcommand\Sources{\Component{Sources}\xspace}
\newcommand\Renderer{\Component{Renderer}\xspace}
\newcommand\Rendering{\Component{Rendering}\xspace}
\newcommand\CellCache{\Component{CellCache}\xspace}


The \MetaData component is responsible for loading meta data information about a dataset from the XML file of the dataset.
The XML is converted into a Java object representation.
For this purpose, we employ the open source SpimData library (available on \href{http://github.com/tpietzsch/spimdata}{\url{http://github.com/tpietzsch/spimdata}}, see also Supplementary Note~\supplFileFormatNumber).
The SpimData object representation is used to create data \Sources that provide input data to the \Renderer.
The meta data provides the number of \setups and \timepoints in the data set, as well as transformations into a common global space (registration of the views).
One \Source is created per \setup. (For a lightsheet microscopy dataset, each combination of channel and acquisition angle is a \setup.)
Each \Source can be queried for image volumes for each \timepoint.
These image volumes are obtained from the \CellCache.

The \CellCache provides infrastructure for loading requested image chunks and maintaining a cache of recently used chunks in RAM.
The \CellCache is instantiated with a loader backend (\eg, \Component{HDF5}) whose the type and parameters are determined from the meta data.
The cache makes use of Java's garbage collection and |SoftReference| mechanism to determine when and which chunks need to be evicted from memory.
Requests to load image chunks are prioritized and enqueued in loading queues.
Low-resolution chunks take precedence over high-resolution chunks, because these load faster and allowing render at least low-resolution images for fluent interactive navigation.
Chunks that are needed to render the current image take precedence over chunks that are predictively loaded (for example to fill in high-resolution data for recently visited locations).
Loading of prioritized chunks is performed asynchronously in one or more background threads.
The \CellCache also takes care of wrapping chunked images as ImgLib2 |CellImg| data containers.
These can be treated by clients as if all data were present in memory, without knowing about the caching mechanism.

The \Renderer consumes a set of \Sources.
As the user navigates the dataset, it request an image volume from each visible \Source for the current \timepoint.
It uses the rendering algorithm described in Supplementary Note~\supplRenderingNumber to show arbitrary slices through the dataset.

\begin{figure}
\centerline{\includegraphics{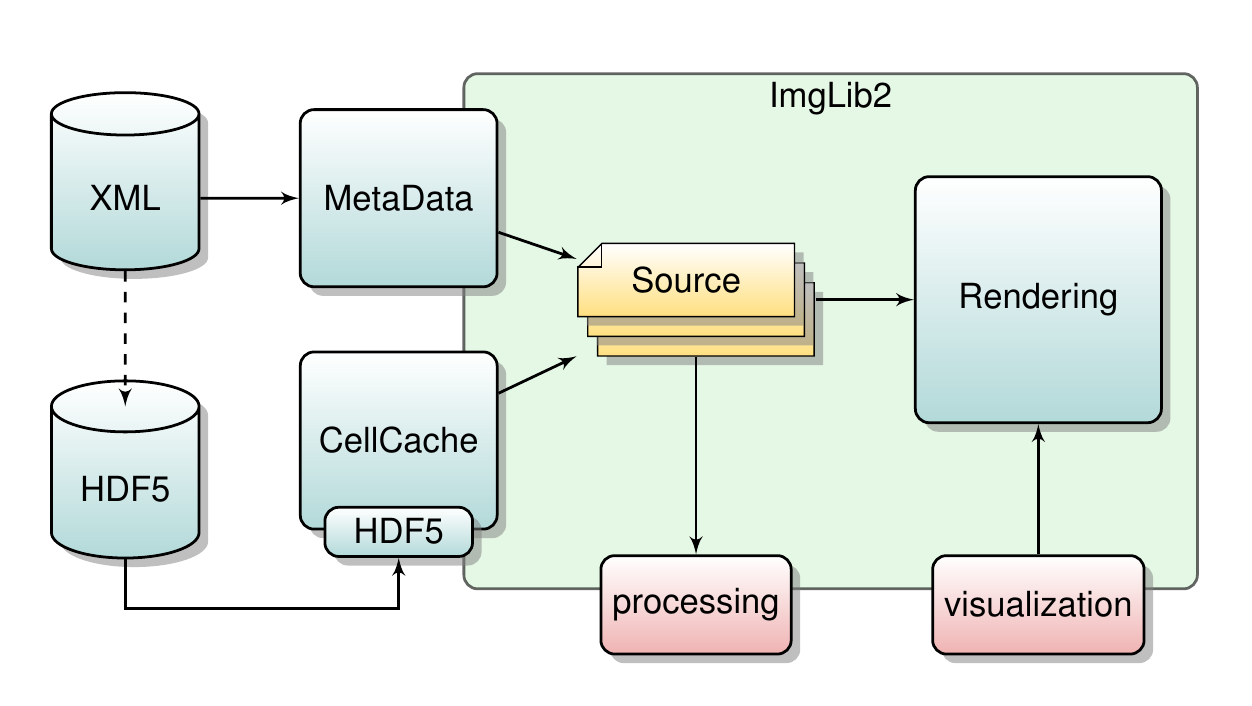}}
\caption{Components of the \bdv architecture and extension points. Data \Sources for \Rendering are created based on \MetaData from XML files and image volumes provided by the \CellCache. Data from \Sources can be accessed for processing.
  Processing results and annotations can be rendered individually or as overlays.}
\label{fig:architecture}
\end{figure}

\begin{figure}
\centerline{\includegraphics{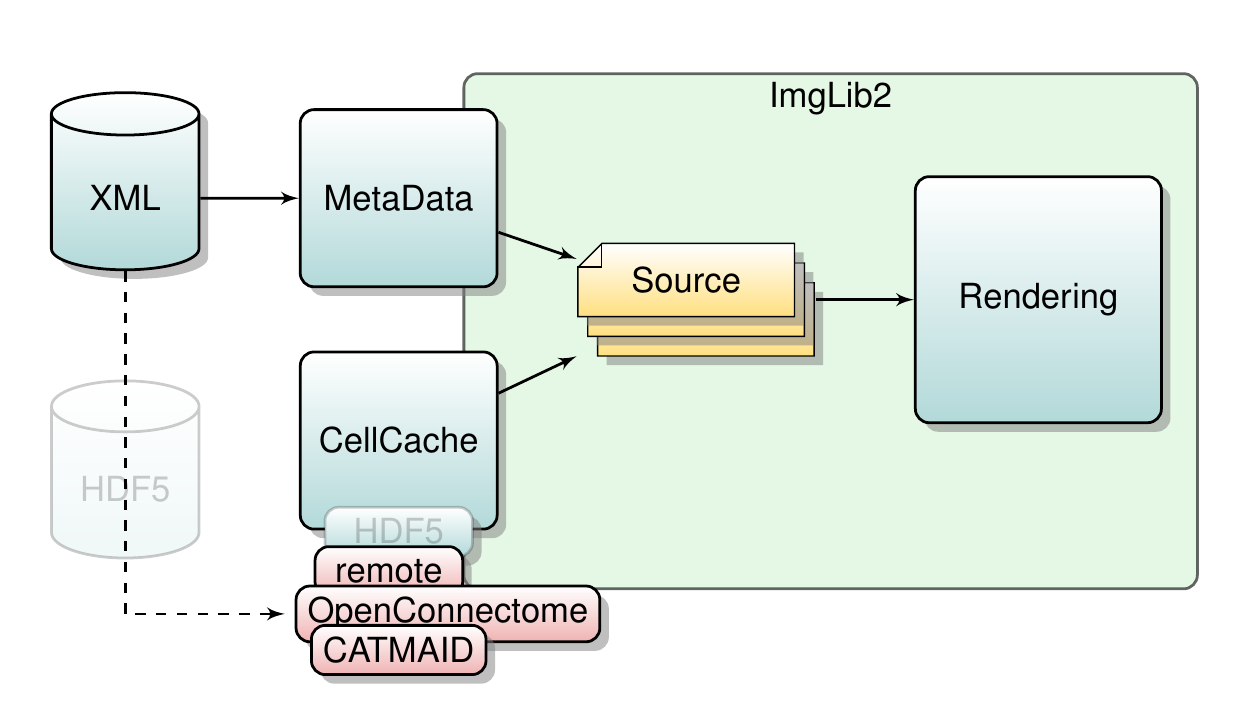}}
\caption{Data Format Extensibility. New data backends can be implemented on top of the cache infrastructure by replacing the HDF5 loader.}
\label{fig:architecture:loader}
\end{figure}

Our architecture provides several extension points (Figure~\ref{fig:architecture} and~\ref{fig:architecture:loader}):
Data provided by the \CellCache can be accessed for processing.
Additional information (annotations, processing results) can be displayed in the \Renderer, blended with the original data.
The central concept to achieve this are \Sources.
A \Source is defined by a leight-weight interface that implements the following functionality.
\begin{itemize}
  \item A method that returns the \emph{voxel type} of the \Source.
    This is the data type of voxels in image volumes provided by the \Source.
    For example, \Sources created from HDF5 files have voxel type |UnsignedShortType|, \ie, volumes with 16-bit precision.
    \Sources created from online CATMAID~\cite{sup:catmaid} resources have voxel type |ARGBType|, \ie, each voxel is a 8-bit ARGB tuple.
  \item A method that returns the number of resolution levels that the \Source provides.
    A simple \Source may provide data for only a single resolution level.
  \item A method that returns whether the \Source provides an image volume for a given \timepoint.
  \item A method that returns an image volume for a given \timepoint and resolution level.
    Volumes may be either bounded and with voxel coordinates on a discrete grid, or
    unbounded with continuously defined coordinates.
    A \Source must provide both flavours
    (however, note that going from one representation to the other using on-the-fly interpolation or rasterization is achieved in a single line of code with ImgLib2).
  \item A method that returns, for a given \timepoint and resolution level, the transformation from the voxel coordinate frame into the global reference frame.
\end{itemize}

In \bdv, \Sources provide data not only for for rendering.
External processing code can access these data, too, by requesting image volumes from \Sources.
These volumes are backed by the \bdv cache, however, for the processing code this is transparent.
It can be assumed that all data are available locally in memory.
Thus, the same unmodified processing code works for large datasets that do not fit the RAM of the local machine and for data that is fetched from remote services, \eg, online resources.

Similaraly, \Sources provide a way to visualize processing results.
Processing results can be wrapped as a \Source and handed to the \Renderer for display, either individual or overlaid with the original data.
Importantly, \Sources provide image volumes as standard ImgLib2 data structures.
This means that voxel access is virtualized, allowing for leight-weight lazily evaluated image volumes.
For further details see Supplementary Note~\ref{sec:renderextensibility} and the example in Figure~\ref{fig:render:continuous}.
Moreover, it is also possible to paint arbitrary 2D overlays on top of the rendered data.

The \bdv architecture and XML file format allow for alternative data backends instead of HDF5 (Figure~\ref{fig:architecture:loader}).
\Bdv already supports alternative backends for remotely served HDF5 datasets, and for the CATMAID~\cite{sup:catmaid} and OpenConnectome~\cite{sup:ocp} web services.
These and any newly implemented data backends can build on the \CellCache infrastructure.
A backend is only required to describe how the data is organized into into chunks, and to provide functionality to load a data chunk into a Java primitive array.\footnote{
  Note, the this data-loading functionality may be arbitrarily simple or complex.
  It can do pre-processing to compensate for limitations of the underlying data source.}
The \CellCache then takes care of deciding which chunks are loaded, how they are cached, and how they are presented as \Sources.

\bibliographystyle{ieeetr}

\end{document}